\documentclass[amsmath,amssymb,aps,10pt,prd,nofootinbib,letterpaper,balancelastpage,notitlepage,superscriptaddress,twocolumn,floatfix,preprintnumbers]{revtex4-2}
\usepackage{graphicx}	
\usepackage{ragged2e}	
\usepackage{bm}		    
\usepackage[sort&compress]{natbib}	 	
	\setcitestyle{square,numbers,comma}	
\usepackage[colorlinks=true,urlcolor=blue,linkcolor=red,citecolor=red]{hyperref}
\usepackage{siunitx}
\usepackage{soul}

\makeatletter 
    
\renewcommand\onecolumngrid{
\do@columngrid{one}{\@ne}%
\def\set@footnotewidth{\onecolumngrid}
\def\footnoterule{\kern-6pt\hrule width 1.5in\kern6pt}%
}

\renewcommand\twocolumngrid{
        \def\footnoterule{
        \dimen@\skip\footins\divide\dimen@\thr@@
        \kern-\dimen@\hrule width.5in\kern\dimen@}
        \do@columngrid{mlt}{\tw@}
}%

\makeatother    

\newcommand{\beq}{\begin{equation}}
\newcommand{\eeq}{\end{equation}}
\newcommand{\bea}{\begin{eqnarray}}
\newcommand{\eea}{\end{eqnarray}}

\newcommand{\gsim}{\lower.7ex\hbox{$\;\stackrel{\textstyle>}{\sim}\;$}}
\newcommand{\lsim}{\lower.7ex\hbox{$\;\stackrel{\textstyle<}{\sim}\;$}}

\newcommand{\be}{\begin{equation}}
\newcommand{\ee}{\end{equation}}
\newcommand{\ba}{\begin{eqnarray}}
\newcommand{\ea}{\end{eqnarray}}

\newcommand{\LL}{\mathcal{L}}
\newcommand{\nl}{\nonumber \\ & \quad }
\newcommand{\phihat}{\bm{\hat\phi}}
\newcommand{\rhat}{\bm{\hat r}}
\newcommand{\rhohat}{\bm{\hat\rho}}
\newcommand{\xhat}{\bm{\hat x}}
\newcommand{\yhat}{\bm{\hat y}}
\newcommand{\zhat}{\bm{\hat z}}
\newcommand{\figref}[2][]{Fig{#1}.~\ref{fig:#2}}		
\newcommand{\secref}[2][]{Sec{#1}.~\ref{sec:#2}}		
\newcommand{\appref}[2][x]{Appendi{#1}~\ref{app:#2}}	
\renewcommand{\eqref}[2][]{Eq{#1}.~(\ref{eq:#2})}		
\newcommand{\citeR}[2][]{Ref{#1}.~\cite{#2}}			

\begin{document}

\title{Maglev for Dark Matter:\\Dark-photon and axion dark matter sensing with levitated superconductors}
\date{\today}
\author{Gerard Higgins}
\email{gerard.higgins@univie.ac.at}
\affiliation{Institute for Quantum Optics and Quantum Information (IQOQI), Austrian Academy of Sciences, A-1090 Vienna, Austria}
\affiliation{Department of Microtechnology and Nanoscience (MC2), Chalmers University of Technology, SE-412 96 Gothenburg, Sweden}
\author{Saarik Kalia}
\email{kalias@umn.edu}
\affiliation{School of Physics \& Astronomy, University of Minnesota, Minneapolis, MN 55455, USA}
\author{Zhen Liu}
\email{zliuphysics@umn.edu}
\affiliation{School of Physics \& Astronomy, University of Minnesota, Minneapolis, MN 55455, USA}

\preprint{UMN-TH-4302/23}
\preprint{FERMILAB-PUB-23-624-SQMS}

\begin{abstract}
Ultraprecise mechanical sensors offer an exciting avenue for testing new physics.  While many of these sensors are tailored to detect inertial forces, magnetically levitated (Maglev) systems are particularly interesting, in that they are also sensitive to electromagnetic forces.  In this work, we propose the use of magnetically levitated superconductors to detect dark-photon and axion dark matter through their couplings to electromagnetism.  Several existing laboratory experiments search for these dark-matter candidates at high frequencies, but few are sensitive to frequencies below $\mathrm{1\,kHz}$ (corresponding to dark-matter masses $m_\mathrm{DM}\lesssim10^{-12}\,\mathrm{eV}$).  As a mechanical resonator, magnetically levitated superconductors are sensitive to lower frequencies, and so can probe parameter space currently unexplored by laboratory experiments.  Dark-photon and axion dark matter can source an oscillating magnetic field that drives the motion of a magnetically levitated superconductor.  This motion is resonantly enhanced when the dark matter Compton frequency matches the levitated superconductor's trapping frequency.  We outline the necessary modifications to make magnetically levitated superconductors sensitive to dark matter, including specifications for both broadband and resonant schemes.  We show that in the $\mathrm{Hz}\lesssim f_\mathrm{DM}\lesssim\mathrm{kHz}$ frequency range our technique can achieve the leading sensitivity amongst laboratory probes of both dark-photon and axion dark matter.
\end{abstract}

\maketitle

\section{Introduction}

Discerning the nature of dark matter (DM) remains one of the major outstanding problems in fundamental physics.  The mass of the particles which constitute DM is largely unconstrained, and so numerous candidates have been proposed over the years, but one class which has garnered increased attention lately is ultralight bosonic DM~\cite{Arias:2012az,kimball2022search}.  This class consists of DM candidates with masses $\lesssim1\,\mathrm{eV}$.  As the local energy density of dark matter has been measured to be $\rho_\mathrm{DM}\approx0.3\,\mathrm{GeV/cm}^3$~\cite{Evans:2018bqy}, these candidates, in turn, have large number densities.
This necessitates that these candidates must be bosonic, and moreover, should behave like classical fields~\cite{lin2018self,Centers:2019dyn}.  Some of the most popular ultralight DM candidates include QCD axions~\cite{Preskill:1982cy,Abbott:1982af,Dine:1982ah}, axionlike particles~\cite{Gra15,co2020predictions}, and dark photons~\cite{Holdom:1986ag,Nelson:2011sf,Graham:2015rva}.  These candidates are particularly intriguing because the QCD axion can solve the strong CP problem~\cite{Peccei:1977hh,Weinberg:1977ma,Wilczek:1977pj}, while axionlike particles and dark photons are predicted by a variety of string compactifications~\cite{cvetivc1996implications,Svrcek:2006yi,Arvanitaki:2009fg}.

These ultralight candidates may possess couplings to electromagnetism~\cite{Holdom:1986ag,Sikivie:1983ip}, and a variety of laboratory experiments have been proposed to search for such couplings~\cite{ehret2010new,Wagner:2010mi,Redondo:2010dp,Horns:2012jf,Betz_2013,Graham:2014sha,Chaudhuri:2014dla,graham2016dark,Caldwell:2016dcw,Anastassopoulos:2017ftl,Baryakhtar:2018doz,armengaud2019physics,Lawson:2019brd,gramolin2021search,Andrianavalomahefa:2020ucg,Gelmini2020,Cantatore:2020obc,Salemi:2021gck,Su:2021jvk,Fedderke:2021iqw,Chiles_2022,haystaccollaboration2023new,romanenko2023new,jiang2023search,sulai2023hunt}.  Many of these experiments search for electromagnetic fields sourced by ultralight DM.  In particular, in the regime where the Compton wavelength of the dark matter $\lambda_\mathrm{DM}$ is much larger than the size of the experiment, the typical signal that these ultralight DM candidates would produce is an oscillating magnetic field~\cite{Chaudhuri:2014dla,Fedderke_2021}.

Various experiments searching for ultralight DM utilize systems which take advantage of resonant enhancements, e.g. lumped-element circuits~\cite{Chaudhuri:2014dla,Salemi:2021gck}, resonant cavities~\cite{Wagner:2010mi,haystaccollaboration2023new}, or layers of dielectric disks~\cite{Caldwell:2016dcw,Baryakhtar:2018doz,Chiles_2022}, in order to increase their sensitivity to DM of a particular Compton frequency (mass).  The frequency range to which each of these experiments is sensitive is set, respectively, by: the inductance and capacitance of the circuit, the size of the cavity, and the spacing between the layers.  It is thus difficult for any of these techniques to probe frequencies below $\mathrm{1\,kHz}$, corresponding to DM masses $m_\mathrm{DM}\lesssim10^{-12}\,\mathrm{eV}$.  In this work, we propose to utilize a mechanical resonator, specifically a magnetically-levitated superconducting particle (SCP), in order to detect the oscillating magnetic field sourced by DM, at frequencies in the Hz to kHz range.

Magnetically levitated superconductors function as ultraprecise accelerometers~\cite{goodkind,Griggs2017,Hofer2023}, and can be employed in a wide range of precision sensing applications. In comparison with optical levitation, magnetostatic levitation allows for the suspension of significantly larger loads~\cite{goodkind}, up to even train-scale objects~\cite{moon2004superconducting}.  Magnetically levitated superconductors have been utilized for gravimetry~\cite{goodkind,Griggs2017}, and have the potential to test quantum physics on macroscopic scales~\cite{RomeroIsart2012,Cirio2012}. The usage of accelerometers to detect $B-L$ dark matter has been actively explored in recent years~\cite{graham2016dark,Carney_2021,Manley:2020mjq, mechanicalqs,Windchime:2022whs,Antypas:2022asj}, and magnetically levitated systems have been proposed as one promising candidate, due to their excellent acceleration sensitivity~\cite{mechanicalqs,Windchime:2022whs,Li:2023wcb,Windchime2023}.
Here, we highlight that magnetically levitated systems are also excellent magnetometers, and as such, can be sensitive to electromagnetically coupled ultralight DM.

The fundamental property underlying the magnetic levitation of a superconductor is its superdiamagnetism, which means that nearly all magnetic fields are expelled out of its interior~\cite{tinkham2004introduction}.  In the presence of an external magnetic field, currents are driven along the surface of the SCP which screen the interior from the magnetic field.  These surface currents then experience a Lorentz force from the external magnetic field, leading to a net force on the SCP (see \figref{scpfig}).  This principle can be used to trap a SCP near the center of an applied static quadrupole field, with trapping frequencies typically in the Hz to kHz range~\cite{GutierrezLatorre2022, GutierrezLatorre2023, Hofer2023}.

The levitation apparatus must be surrounded by magnetic shielding in order to isolate the SCP from environmental fields.  Inside this shield, ultralight DM can source an oscillating magnetic field signal, similar to the one sourced in experiments like DM Radio~\cite{Chaudhuri:2014dla}.  If the apparatus is positioned off-center within the shield, this signal can be nonzero in the vicinity of the apparatus.  This additional field can then perturb the equilibrium position of the SCP, leading to oscillatory motion of the particle.  If the frequency of this oscillation, which is set by the DM mass, matches the trapping frequency, then the motion will be resonantly enhanced.  Magnetically levitated SCPs thus provide an excellent context in which to resonantly search for electromagnetic couplings of DM with $m_\mathrm{DM}\lesssim10^{-12}\,\mathrm{eV}$.  In this work, we will explore both resonant and broadband detection schemes to search for ultralight DM in this mass range.

This work is structured as follows.  In \secref{levitation}, we outline how a superconductor can be magnetically levitated.  We discuss the physics of trapping a SCP, as well as the possible readout schemes and potential range of system parameters.  In \secref{DM}, we review the physics of the ultralight DM candidates considered in this work. 
 These include dark-photon dark matter (DPDM) and axion DM.%
\footnote{Throughout this work, we simply use ``axion" to refer to both the QCD axion and axionlike particles.}
In \secref{sensitivity}, we discuss the relevant noise sources for our setup and project sensitivities to both DM candidates.  We consider both a broadband scheme using a single experiment and a scanning scheme using several resonant experiments, and outline the parameter choices relevant to each of these schemes.  Finally, in \secref{discussion}, we discuss our results and possible future improvements.  We also perform detailed computations in our appendices.  In \appref{response}, we derive the response of a spherical SCP to an applied magnetic field.  In \appref{axion}, we derive the axion DM magnetic field signal sourced inside a rectilinear magnetic shield.  We make all the code used in this work publicly available on Github~\cite{github}.

\section{Levitated Superconductors}
\label{sec:levitation}

In this section, we discuss the magnetic levitation of a SCP.  First, we show how a SCP can be trapped near the center of a static quadrupolar magnetic field.  Next, we discuss various methods of reading out the motion of a SCP inside the trap.  Finally, we outline the physical limitations of the setup, and the resulting range of parameters that can be achieved with such a system.

\subsection{Trapping}
\label{sec:trapping}

\begin{figure*}[t]
\includegraphics[width=\textwidth]{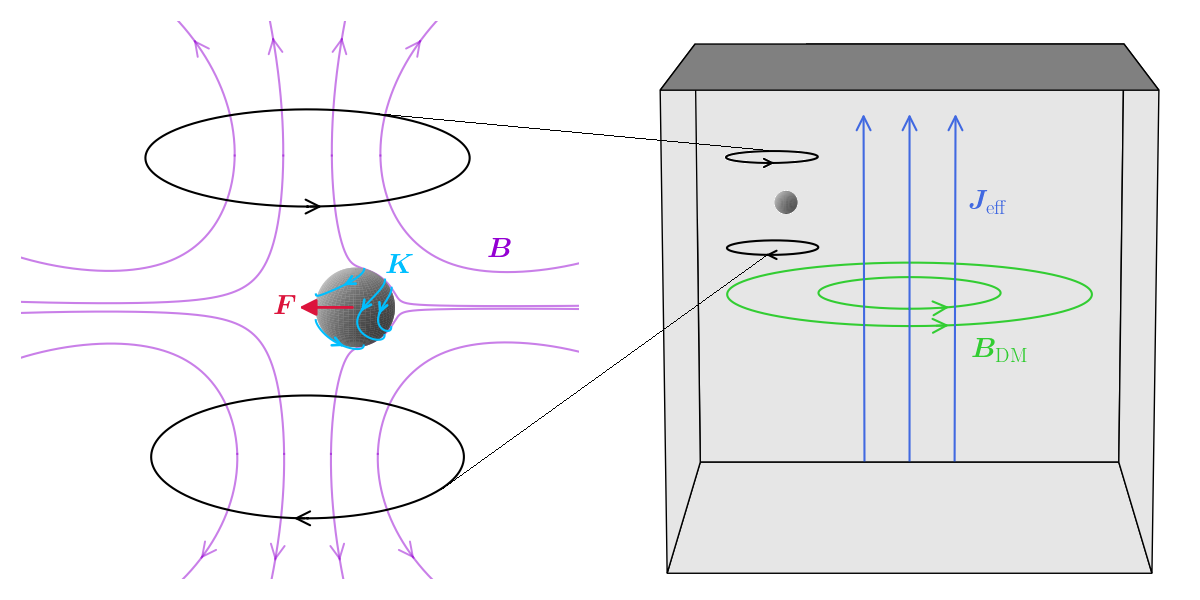}
\caption{\label{fig:scpfig}%
    Magnetic levitation of a SCP.  The levitation apparatus (shown on the left) consists of two current-carrying coils arranged in an anti-Helmholtz-like configuration, i.e. carrying currents in opposite directions.  Together these coils source a quadrupole magnetic field (shown in purple), which can trap a SCP.  If the SCP is displaced from the center of the trap (the point at which $\bm B=0$), surface currents (shown in light blue) will run on the SCP to screen the magnetic field out of its interior.  These surface currents then experience a Lorentz force in the presence of the magnetic field, leading to a net restoring force (shown in red) which drives the SCP back to the center of the trap.  The trap is typically located within a magnetic shield (shown on the right).  Inside of this shield, ultralight DM can be parametrized by an effective current (shown in dark blue), which sources an oscillating magnetic field signal (shown in green).  In the DPDM case, the direction of the effective current is given by the DPDM polarization.  In the axion case, it is given by the quadrupole magnetic field trap.  The DM-induced magnetic field can displace the equilibrium position of the trap, resulting in oscillatory motion of the SCP.  Note that since this magnetic field signal vanishes at the center of the shield, the trap must be located off-center within the shield in order to be sensitive to DPDM or axion DM.}
\end{figure*}

The magnetic trap is formed by a quadrupolar field, which confines the SCP near its center, at the point where the magnetic field vanishes.  Such a magnetic field can be created by two coils carrying currents in opposite directions, also known as an anti-Helmholtz-like configuration (see \figref{scpfig}).  To understand the effect of the trap on the SCP, let us expand the magnetic field in the vicinity of the trap center to linear order as
\begin{equation}
    B_i(\bm x,t)=B_{0,i}(t)+b_{ij}(t)x_j,
    \label{eq:Bfield}
\end{equation}
where the Einstein summation convention is implicit in the second term.  Here $\bm B_0$ represents the magnetic field at the center of the coordinate system, while $b_{ij}$ describes the magnetic field gradients near the center.  (Note that Gauss's law of magnetism enforces $\sum_ib_{ii}=0$).

In the absence of beyond-the-Standard-Model effects, the only contribution to $\bm B$ is the applied quadrupole trap, which is static.  Because of this, we can choose a coordinate system in which the magnetic field vanishes at the origin, i.e. $\bm B_0(t)=0$, and in which $b_{ij}$ is diagonal.  Then \eqref{Bfield} simplifies to
\begin{equation}
    \bm B_\mathrm{trap}(\bm x,t)=b_{xx}x\xhat+b_{yy}y\yhat+b_{zz}z\zhat.
    \label{eq:Bfield_quad}
\end{equation}
When we introduce a DM signal, the total magnetic field will not take this simple form, as $\bm B$ will exhibit a time dependence.

A superconducting sphere of volume $V$, located at position $\bm x$ within the magnetic field in \eqref{Bfield}, will experience a force%
\footnote{Throughout, we use natural units $\hbar=c=k_B=\mu_0=1$.}
\begin{equation}
    F_i(\bm x,t)=-\frac32Vb_{ji}B_j(\bm x,t)
    \label{eq:sc_force}
\end{equation}
(see \appref{response} or \citeR[s]{Hofer_2019,Hofer2023} for derivation).  Microscopically, this force occurs because the local magnetic field drives surface currents on the SCP, in order to screen the magnetic field out of its interior.  These currents then experience a Lorentz force in the presence of the magnetic field (see \figref{scpfig}).  Note that because the net force is given by the difference between the Lorentz forces on either side of the sphere, \eqref{sc_force} depends not only on the magnetic field, but also on its gradient $b_{ij}$ across the sphere.

This force can alternatively be understood by rewriting \eqref{sc_force} as $\bm F=-\nabla U$, where
\begin{equation}
    U=\frac34V|\bm B|^2.
\end{equation}
Heuristically, this potential can be interpreted as the amount of energy that it takes for the superconducting sphere to screen out the local magnetic field.  The sphere will therefore settle at the point of lowest total magnetic field.  In the case of a static quadrupole field, this will be the center of the trap $\bm x=0$.

This can be seen even more directly by plugging \eqref{Bfield_quad} into \eqref{sc_force} to find
\begin{equation}
    \bm F_\mathrm{trap}(\bm x,t)=-\frac32V\left(b_{xx}^2x\xhat+b_{yy}^2y\yhat+b_{zz}^2z\zhat\right).
    \label{eq:Fquad}
\end{equation}
This expression makes it clear that the trap creates a restoring force towards $\bm x=0$, so that the system acts as a harmonic oscillator.  The resonant frequencies of the trap are simply given by~\cite{Hofer2023}
\begin{equation}
    f_i=\sqrt{\frac 3{8\pi^2\rho}} \, b_{ii},
    \label{eq:trap_freq}
\end{equation}
where $\rho$ is the density of the sphere.  As we will see in \secref{DM}, the magnetic field signal induced by ultralight DM can drive this harmonic oscillator.  If the frequency of the driving signal (which is set by the ultralight DM mass) matches one of the trapping frequencies in \eqref{trap_freq}, then the oscillator will ring up resonantly.

\subsection{Readout}

The motion of the levitated SCP can be read out in different fashions.  One method relies on placing a pickup coil close to the particle.  As the SCP moves, it distorts the magnetic trapping field, causing the magnetic flux threading the pickup coil to change.  This flux can be transferred to a sensitive magnetometer, such as a SQUID~\cite{RomeroIsart2012, Hofer2023, GutierrezLatorre2023} or a SQUID coupled to a microwave resonator~\cite{Schmidt2024}, which outputs a signal describing the particle motion.

Another method also makes use of a pickup coil close to the particle, but uses a different mechanism for sensing the particle motion.  As the SCP moves, it changes the inductance of the pickup coil, due to the SCP's superdiamagnetism.  This inductance change can be measured to probe the particle motion~\cite{Griggs2017,Chan1987a}.

Alternatively, the particle motion can be measured using optical interferometry~\cite{Hansen2024}.  Specifically, one can form a Michelson interferometer, with a reflective SCP acting as the mirror at the end of one of the interferometer arms.  In principle, each of these methods allows the particle motion to be probed close to the standard quantum limit (SQL).  In this work, we will primarily consider the SQUID readout.  The sensitivity of this readout scheme will be discussed further in \secref{sensitivity}.

\subsection{Range of system parameters}
\label{sec:range}

Here we discuss physical limitations of this levitation geometry, which set the viable range of parameters that can be achieved.  First, \eqref{trap_freq} implies that the frequency range of our setup is constrained by the range of achievable magnetic field gradients and particle densities, namely%
\footnote{Throughout the rest of this work, we write $f_0=\omega_0/2\pi$ and $b_0$, rather than $f_i$ and $b_{ii}$, to refer to the trapping frequency and magnetic field gradient, in contexts where we are agnostic about which mode is being excited.  These quantities are still related by \eqref{trap_freq}.}
\begin{equation}
    f_0\sim170\,\mathrm{Hz}\cdot\sqrt{\frac{0.1\,\mathrm{g/cm^3}}\rho}\left(\frac{b_0}{10\,\mathrm{T/m}}\right).
\end{equation}
Densities of $0.1\,\mathrm{g/cm^3}$ can be achieved by using a hollow SCP.  A SCP of mass 1\,g and density $0.1\,\mathrm{g/cm^3}$ would require a thickness of $\sim50\,\mu\mathrm{m}$ (see \citeR{goodkind} for levitation of similarly sized hollow SCPs). Such a particle is around $3\,\mathrm{cm}$ across. Field gradients of up to $\sim100\,\mathrm{T/m}$ have been produced in cm-scale traps \cite{Hofer2023}, so we find it reasonable to consider trapping frequencies $f_0\lesssim100\,\mathrm{Hz}$.

Additionally, for sufficiently low trapping frequencies, gravity can displace the vertical equilibrium position of the SCP.  By balancing the force of gravity $\bm F_g=-mg\zhat$ with \eqref{Fquad}, we see that the vertical displacement of the equilibrium will be
\begin{equation}
    \Delta z=\frac g{4\pi^2f_z^2}\sim3\,\mathrm{cm}\cdot\left(\frac{3\,\mathrm{Hz}}{f_z}\right)^2.
    \label{eq:grav_displacement}
\end{equation}
To avoid significant displacements from gravity, in this work, we will focus on the range of trapping frequencies $3\,\mathrm{Hz}\lesssim f_0\lesssim100\,\mathrm{Hz}$.

The size of the SCP is also constrained by the critical fields of the superconducting material out of which it is made.
As the size of the SCP is increased, the magnetic field strength at its surface will increase (due to the magnetic field gradients $b_{ii}$), and so its superconductivity can be broken if the SCP is too large.  When the SCP is located at the center of the trap, the maximum magnetic field strength on its surface is given by
\begin{align}
    B_\mathrm{max}&\sim b_0\mathcal R\\
    &\sim80\,\mathrm{mT}\cdot\left(\frac m{1\,\mathrm g}\right)^{1/3}\left(\frac\rho{0.1\,\mathrm{g/cm^3}}\right)^{1/6}\left(\frac{f_0}{100\,\mathrm{Hz}}\right),
    \label{eq:Bmax}
\end{align}
where $\mathcal R$ is the characteristic length of the SCP.  Typical type-I superconducting materials, such as Pb and Ta, have critical field strengths of up to 80\,mT~\cite{Shaw1960, Chanin1972}, so in this work we restrict ourselves to SCPs no larger than $m=1\,\mathrm g$.  We note, however, that thin films of TiN have been shown to have critical field strengths of up to 5\,T~\cite{Pracht2012}, so larger SCPs may be possible.

Finally, as this system acts as a harmonic oscillator, it exhibits a characteristic dissipation rate $\gamma$.  We anticipate the main source of dissipation to be gas collisions with the SCP.  The dissipation rate from gas collisions is given by~\cite{Carney_2021,Hinkle}
\begin{align}
    \gamma\sim\frac{PA}{m\bar v_\mathrm{gas}}&\sim2\pi\cdot10^{-8}\,\mathrm{Hz}\cdot\left(\frac P{10^{-7}\,\mathrm{Pa}}\right)\left(\frac{1\,\mathrm g}m\right)^{1/3}\nl
    \cdot\left(\frac{0.1\,\mathrm{g/cm}^3}\rho\right)^{2/3}\sqrt{\left(\frac{m_\mathrm{gas}}{4\,\mathrm{Da}}\right)\left(\frac{10\,\mathrm{mK}}T\right)},
\end{align}
where $P$ is the gas pressure, $A$ is the cross-sectional area of the SCP, and $\bar v_\mathrm{gas}\sim\sqrt{T/m_\mathrm{gas}}$ is the mean velocity of the gas molecules (which have mass $m_\mathrm{gas}$).  Other potential sources of dissipation include flux creep and eddy current damping.  Flux creep is the movement of unpinned flux lines within the SCP~\cite{Hofer2023, Brandt1988}.  Flux pinning occurs in type-II superconductors, and so flux creep can be eliminated by using a SCP made from a type-I superconducting material with few crystalline domains.  Eddy current damping occurs when the motion of the SCP causes magnetic field changes which drive currents in nearby resistive conductors with nonzero resistance.  This dissipation can be mitigated by surrounding the levitation apparatus by a superconducting shield (see \figref{scpfig}), and ensuring all materials inside the shield are either superconductors or electrical insulators.  We therefore expect $\gamma\sim2\pi\cdot10^{-8}\,\mathrm{Hz}$ to be an achievable benchmark for the dissipation rate.%
\footnote{Even lower dissipation rates can be achieved with lower pressures.  However, as we will see, decreasing the dissipation rate further will not necessarily improve our sensitivity.  In our ``broadband" setup, thermal noise is subdominant so that $\gamma$ becomes irrelevant.  In the ``scanning" setup, smaller $\gamma$ will improve the sensitivity on resonance but decrease the linewidth of each individual experiment [see \eqref{width}].  It will then require a longer total integration time in order to scan the same frequency range.  A dedicated analysis of the scanning strategy would be required to determine how to take advantage of a lower dissipation rate in the scanning setup.}

\section{Dark matter signals}
\label{sec:DM}

In this section, we review two ultralight DM candidates, dark-photon dark matter (DPDM) and axion DM, and derive the signals that they can effect on a levitated SCP through their coupling to electromagnetism.  As we will see, both DM candidates can be described by an effective current.  Within the confines of the magnetic shield surrounding the levitation setup, this effective current sources an oscillating magnetic field signal, just as inside shielded experiments like DM Radio~\cite{Chaudhuri:2014dla}.  This magnetic field will then drive oscillatory motion of the SCP.

\subsection{Dark-photon dark matter}

A kinetically mixed dark photon $A'$ of mass $m_{A'}$ and kinetic mixing parameter $\varepsilon$ is described by the Lagrangian%
\footnote{The Lagrangian for the mixed photon--dark-photon system can be written in multiple different bases (see Sec.~II\,A and Appendix~A of \citeR{Fedderke_2021} for a detailed review).  In this work, we operate only in the so-called ``interaction basis," in which the Lagrangian is given by \eqref{DPDM_Lagrangian}.  In this basis, only $A$ interacts with SM currents at leading order.  However, $A$ and $A'$ are not propagation eigenstates, and so will mix as they propagate through vacuum.}
\begin{align}
    \LL_{A'}&\supset-\frac14F_{\mu\nu}F^{\mu\nu}-\frac14F'_{\mu\nu}F'^{\mu\nu}+\frac12m_{A'}^2A'_\mu A'^\mu\nl
    +\varepsilon m_{A'}^2A_\mu A'^\mu-J_\mathrm{EM}^\mu A_\mu,
    \label{eq:DPDM_Lagrangian}
\end{align}
where $F'_{\mu\nu}=\partial_\mu A'_\nu-\partial_\nu A'_\mu$ is the field-strength tensor for the dark photon, and $J^\mu_\mathrm{EM}$ is the Standard Model electromagnetic current.  By comparing the last two terms in \eqref{DPDM_Lagrangian}, we can see that $A'$ has a similar effect to a current.  In particular, if we take $\varepsilon\ll1$ so that there is negligible backreaction on $A'$ and consider the limit where the DPDM is non-relativistic $v_\mathrm{DM}\sim10^{-3}\ll1$, then the only effect $A'$ has on electromagnetism is to modify the Amp\`ere-Maxwell law by~\cite{Fedderke_2021}%
\footnote{Throughout, we use unbolded symbols $A'$ to denote four-vectors and bolded symbols $\bm A'$ to denote three-vectors.}
\begin{equation}
    \nabla\times\bm B-\partial_t\bm E=\bm J_\mathrm{eff},
    \label{eq:ampere}
\end{equation}
where
\begin{equation}
    \bm J_\mathrm{eff}=-\varepsilon m^2\bm A'
    \label{eq:DPDM_current}
\end{equation}
is the ``effective current" induced by the DPDM.

Naively, \eqref{ampere} implies that the DPDM may generate either an electric or magnetic field.  A well-controlled magnetic levitation setup must however occur inside some magnetic shielding (see \figref{scpfig}).  This magnetic shield typically acts a perfect conductor, and so the tangential electric field at its surface must vanish.  The DM-induced signal will have a wavelength matching the Compton wavelength of the DM, $\lambda_\mathrm{DM}\gtrsim10^7\,\mathrm m$ (for $f_\mathrm{DM}\lesssim100\,\mathrm{Hz}$).  This wavelength sets the length scale on which the electric field can vary, and will be much larger than the characteristic size of the shielding.  Therefore, since the tangential electric field vanishes at the walls of the shield, it will typically be small everywhere inside the shield.  In other words, the dominate signal of DPDM inside the shield will typically be a \emph{magnetic} field (see \citeR[s]{Chaudhuri:2014dla,Fedderke_2021} for similar discussion and examples).  Because the electric field can be neglected, this magnetic field signal should satisfy%
\footnote{Note that the $\bm B$ predicted by \eqref{ampere_approx} is the observable magnetic field associated with $A$, not the dark magnetic field associated with $A'$.  While the latter is suppressed by $v_\mathrm{DM}$, $\bm B$ need not be.}
\begin{equation}
    \nabla\times\bm B\approx\bm J_\mathrm{eff}.
    \label{eq:ampere_approx}
\end{equation}

As an example, let us consider the case where the shield is a cylinder of radius $L$ (and arbitrary height).  Suppose that the DPDM is polarized along the axis of the cylinder, which we will identify with the $z$-axis. That is, in the non-relativistic limit, the spatial components of $A'$ are given by
\begin{equation}
    \bm A'(\bm x,t)= A'_0\cos(m_{A'}t)\zhat,
    \label{eq:DPDM_wave}
\end{equation}
(and the temporal component of $A'$ is suppressed by $v_\mathrm{DM}$).  This corresponds to an effective current, given by \eqref{DPDM_current}.  If $m_{A'}L\ll1$, then \eqref{ampere_approx} applies, and solving it yields the magnetic field signal~\cite{Chaudhuri:2014dla,Fedderke_2021}%
\footnote{\label{ftnt:norm}%
The DPDM amplitude is normalized by $\frac12m_{A'}^2\langle|\bm A'|^2\rangle=\rho_\mathrm{DM}\approx0.3\,\mathrm{GeV/cm}^3$, where the average $\langle\cdots\rangle$ is taken over many coherence times (the timescale over which the amplitude in \eqref{DPDM_wave} varies; see discussion in \secref{sensitivity}).  Generically, $\bm A'$ can point in any direction, but will have some nonzero projection onto the $z$-axis.  Therefore in this estimate and in the DPDM sensitivity in \figref{sensitivity}, we take $m_{A'}A'_0\sim\sqrt{\frac{2\rho_\mathrm{DM}}3}$.  The estimate in \eqref{axion_signal_est} and the axion sensitivity in \figref{sensitivity}, on the other hand, take $m_aa_0\sim\sqrt{2\rho_\mathrm{DM}}$, since the axion DM has no inherent direction.}
\begin{align}
    \label{eq:DPDM_signal}
    \bm B_{A'}(\bm x,t)&=-\frac12\varepsilon m_{A'}^2A'_0r\cos(m_{A'}t)\phihat\\
    &\sim5\times10^{-20}\,\mathrm{T}\left(\frac\varepsilon{10^{-8}}\right)\left(\frac{f_{A'}}{100\,\mathrm{Hz}}\right)\left(\frac r{1\,\mathrm m}\right),
    \label{eq:DPDM_signal_est}
\end{align}
where $r$ denotes the distance from the axis of the cylindrical shield, and $\phihat$ denotes the azimuthal direction.  Note that $B_{A'}$ vanishes at the center of the cylindrical shield $r=0$.  Therefore, in order to be sensitive to the DPDM signal, it will be important that the magnetic levitation setup is positioned \emph{off-center} within the magnetic shield.

The total field that the magnetically levitated particle experiences will be a combination of the static quadrupole trap and the oscillating DPDM signal.  In other words, \eqref{Bfield} will consist of the terms in \eqref{Bfield_quad}, along with an additional (time-dependent) contribution from the DPDM signal given by \eqref{DPDM_signal}.  As the quadrupole gradient $b_{ij}$ is much larger than the gradient of \eqref{DPDM_signal}, the second term in \eqref{Bfield} will receive negligible corrections. In particular, this implies that the trapping frequencies will remain unchanged.

Instead, the dominant effect of the DPDM signal in \eqref{DPDM_signal} will be to give a time-dependent contribution to the first term in \eqref{Bfield}.  Concretely, let us choose coordinates similar to those used in \eqref{Bfield_quad}, i.e. let $\bm x=0$ denote the point for which the \emph{time-averaged} magnetic field vanishes, $\langle\bm B_0(t)\rangle=0$.  Moreover, we can take coordinates where $b_{ij}$ is diagonal.  Let us suppose the trap is oriented so that one of these coordinate directions is the $z$-direction (the axial direction of the shield).  Then if the center of the trap $\bm x=0$ is displaced by a distance $r$ along the $x$-direction from the axis of the shield, the total magnetic field in the vicinity of the trap center will be
\begin{equation}
    \bm B(\bm x,t)=\bm B_\mathrm{trap}(\bm x,t)-\frac12\varepsilon m_{A'}^2A'_0r\cos(m_{A'}t)\yhat,
\end{equation}
where $\bm B_\mathrm{trap}$ is as in \eqref{Bfield_quad}.  Plugging this into \eqref{sc_force}, we find that the SCP experiences a force
\begin{equation}
    \bm F(\bm x,t)=\bm F_\mathrm{trap}(\bm x,t)+\frac34\varepsilon m_{A'}^2A'_0\cdot Vb_{yy}r\cdot\cos(m_{A'}t)\yhat,
    \label{eq:DPDM_force}
\end{equation}
where $\bm F_\mathrm{trap}$ is the restoring force from \eqref{Fquad}.  The second term represents a driving force, which will drive oscillatory motion along the $y$-direction.  If $m_{A'}\approx2\pi f_y$, this translational mode will be resonantly driven.

\subsection{Axion dark matter}

Levitated SCPs may also be sensitive to axion DM which couples to photons.  An axionlike particle $a$, with mass $m_a$ and coupling $g_{a\gamma}$ to photons, is described by the Lagrangian
\begin{equation}
    \LL_a\supset\frac12\partial_\mu a\partial^\mu a-\frac14F_{\mu\nu}F^{\mu\nu}-\frac12m_aa^2+\frac14g_{a\gamma}aF_{\mu\nu}\tilde F^{\mu\nu},
    \label{eq:axion_lagrangian}
\end{equation}
where $\tilde F^{\mu\nu}=\frac12\epsilon^{\mu\nu\rho\sigma}F_{\rho\sigma}$.  In the non-relativistic limit, the axion DM is uniform in space and oscillates at its Compton frequency (corresponding to its mass $m_a$), i.e. it takes the form
\begin{equation}
    a(\bm x,t)=a_0\cos(m_at).
    \label{eq:axion_wave}
\end{equation}
Much like in the case of DPDM, in the non-relativistic limit, the only effect of the last term in \eqref{axion_lagrangian} is to add an effective current to the Amp\`ere-Maxwell law, as in \eqref{ampere}.  In the axion case, this current takes the form~\cite{Arza_2022,Sikivie:1983ip,Wilczek:1987edt}
\begin{equation}
    \bm J_\mathrm{eff}=-g_{a\gamma}(\partial_ta)\bm B.
    \label{eq:axion_current}
\end{equation}
One important difference from the DPDM case is that an applied magnetic field is required in order for the axion to convert into an electromagnetic signal [as can be seen from the presence of $\bm B$ in \eqref{axion_current}]. Conveniently, in our case, the quadrupole trap itself can act as the necessary applied magnetic field!

As in the DPDM case, this current should produce an oscillating magnetic field inside the shield.  However, in the axion case, the magnetic field response is much more difficult to compute.  From \eqref{axion_current}, we see that in the axion case, the direction of the effective current is set by the static magnetic field.  Therefore the effective current, in this case, will inherit the complicated shape of the trapping field (which depends on how exactly the trap is implemented).  Moreover, just as in the DPDM case, the trap must be positioned off-center within the shield, otherwise the magnetic field sourced by $\bm J_\mathrm{eff}$ will vanish at the center of the trap, by symmetry (see \appref{axion}).  The computation thus amounts to determining the response of a cavity to a complicated asymmetric current distribution.

In \appref{axion}, we compute the signal in the case where the shield is rectilinear and the trap is created by two coils in an anti-Helmholtz-like configuration.  The exact signal must be computed numerically, but we can derive a parametric estimate analytically, in terms of the dimensions of the shield $L$, the radius of the coils $R$, and the distance between the coils $2h$.  We find that the axion-induced magnetic field response at the center of the trap should be
\begin{align}
    \label{eq:axion_signal}
    B_a(0,t)&\sim\mathcal O(0.1)\cdot\frac{g_{a\gamma}m_aa_0b_0\left(R^2+h^2\right)^{5/2}}{L^3}\sin(m_at)\\
    &\sim3\times10^{-20}\,\mathrm{T}\left(\frac{g_{a\gamma}}{10^{-10}\,\mathrm{GeV}^{-1}}\right)\left(\frac {f_0}{100\,\mathrm{Hz}}\right)\nl\cdot\sqrt{\frac\rho{0.1\,\mathrm{g/cm}^3}}\left(\frac h{10\,\mathrm{cm}}\right)^5\left(\frac{100\,\mathrm{cm}}L\right)^3,
    \label{eq:axion_signal_est}
\end{align}
where we have taken $h\sim R$.  The constant of proportionality  in \eqref{axion_signal} depends on the exact position of the trap within the cavity (and as mentioned above, will be zero if the trap is positioned in a sufficiently symmetric location).  As in the DPDM case, this magnetic field will drive the oscillatory motion of the SCP.

\section{Sensitivity}
\label{sec:sensitivity}

In this section, we derive the sensitivity of levitated SCPs to ultralight DM.  To do so, we must first discuss the relevant noise sources.  This section discusses three primary sources: thermal noise, measurement imprecision noise, and measurement backaction noise.  The latter two of these depend on the readout scheme that is used.  This work considers a SQUID readout, although similar noise sources exist for other readout schemes.  Once we have enumerated the noise sources, we discuss the trade-off between imprecision and backaction noise, controlled by the coupling strength of the readout scheme.  We outline two possible choices in this trade-off, one corresponding to a broadband detection scheme and one corresponding to a resonant detection scheme.  Finally, we estimate the sensitivity of both these schemes to DPDM and axion DM.

\subsection{Noise sources}

The first relevant noise source in our system is thermal noise.  By the fluctuation-dissipation theorem, the thermal force noise acting on the SCP is given by $S_{FF}^\mathrm{th}=4m\gamma T$, where $m$ is the mass of the SCP, and $\gamma$ and $T$ are the dissipation rate and temperature of the system~\cite{RKubo_1966}.  To compare with \eqref[s]{DPDM_signal_est} and (\ref{eq:axion_signal_est}), it will be useful to translate this into a noise power spectral density (PSD) for the magnetic field [via \eqref{sc_force}]
\begin{align}
    \label{eq:thermal_noise}
    S_{BB}^\mathrm{th}&=\frac{16m\gamma T}{9V^2b^2}=\frac{8\rho\gamma T}{3m\omega_0^2}\\&\sim7\times10^{-39}\,\mathrm{T^2/Hz}\left(\frac{1\,\mathrm g}m\right)\left(\frac\rho{0.1\,\mathrm{g/cm^3}}\right)\nl\cdot\left(\frac\gamma{2\pi\cdot10^{-8}\,\mathrm{Hz}}\right)\left(\frac T{10\,\mathrm{mK}}\right)\left(\frac{100\,\mathrm{Hz}}{f_0}\right)^2.
    \label{eq:thermal_noise_est}
\end{align}

The second noise source of interest is imprecision noise.  As mentioned above, the details of this noise source depend on the readout scheme used.  Here, we consider a SQUID readout, in which case imprecision noise arises from flux noise within the SQUID.  The DM-induced magnetic field exerts a force on the SCP, causing it to move and distort the local magnetic field.  This, in turn, changes the flux measured by the SQUID.  Conversely, uncertainty in the measured flux of the SQUID results in uncertainty in the DM-induced magnetic field.  Let us denote the internal flux noise of the SQUID by $S_{\phi\phi}(\omega)$.  We can parameterize the coupling between the position of the SCP and the measured flux of the SQUID by a parameter $\eta$, which can be varied, e.g., by changing the inductance of the pickup coil or its position relative to the SCP~\cite{Hofer2023}.  The flux noise of the SQUID is then related to noise in the position of the SCP via $S_{xx}^\mathrm{imp}=S_{\phi\phi}/\eta^2$.  We can convert this position noise into a magnetic field noise PSD (as a function of frequency $\omega$) via
\begin{align}
    S_{BB}^\mathrm{imp}(\omega)&=\frac{4S_{FF}^\mathrm{imp}}{9V^2b^2}=\frac{4S_{xx}^\mathrm{imp}(\omega)}{9V^2b^2|\chi(\omega)|^2}\\
    &=\frac{2\rho S_{\phi\phi}(\omega)}{3m^2\omega_0^2\eta^2|\chi(\omega)|^2},
    \label{eq:measurement_noise}
\end{align}
where
\begin{equation}
    \chi(\omega)=\frac1{m(\omega_0^2-\omega^2-i\gamma\omega)}
\end{equation}
denotes the mechanical susceptibility.  Note that while the thermal noise in \eqref{thermal_noise} is frequency-independent, the imprecision noise in \eqref{measurement_noise} does depend on frequency.  In particular, the imprecision noise becomes significantly suppressed at the trapping frequency $\omega=\omega_0$.

The final relevant source of noise is back-action noise.  This arises from current noise $S_{JJ}(\omega)$ within the SQUID.  A current $J$ circulating in the SQUID will generate local magnetic fields which back-react on the SCP with a force $-\eta J$~\cite{Hofer2023}.  The larger the coupling $\eta$ is, the stronger the back-reaction on the SCP will be.  Therefore, when choosing $\eta$, there exists a trade-off between imprecision noise and back-action noise.  The magnetic field noise PSD associated with back-action noise is given by
\begin{equation}
    S_{BB}^\mathrm{back}(\omega)=\frac{2\rho\eta^2S_{JJ}(\omega)}{3m^2\omega_0^2}.
\end{equation}
As with thermal noise, back-action noise is frequency-independent (up to any frequency dependence coming from $S_{JJ}$; see next section).

We also note one additional noise source, namely vibrational noise.  External vibrations of the system lead to position noise $S_{xx}^\mathrm{vib}$, and as in the case of imprecision noise, this will manifest as noise in the force and magnetic field.  Vibrational noise is, however, not inherent to the readout scheme, and can be mitigated by various means.  As in \citeR{Hofer2023}, the experimental apparatus can be hung from a vibration isolation system to reduce vibrational noise.  Further, instead of utilizing just a single levitation apparatus, a second copy can be set up at the center of the same shield.  Then both copies will experience the same external vibrations, while only the first will be sensitive to the DM signal.  The relative displacement of the two sensors can then be used to isolate the DM signal from external vibrations.  We leave a more detailed study of vibrational noise to future work.

\subsection{Choice of the coupling $\eta$}

Before we can estimate the size of the imprecision and back-action noise sources, we must decide on an appropriate choice for the coupling $\eta$.  First, let us observe that $S_{\phi\phi}$ and $S_{JJ}$ are described by an uncertainty relation $\sqrt{S_{\phi\phi}S_{JJ}}=\kappa$, where $\kappa\geq1$ is referred to as the SQUID's energy resolution~\cite{Voss1981}.  The limiting case $\kappa=1$ corresponds to the SQL.  State-of-the-art SQUIDs can achieve $\kappa\approx5$ \cite{Awschalom1988, Carelli1998, Vasyukov2013}.  We note that SQUIDs typically display $1/f$ noise at frequencies $\lesssim10\,\mathrm{kHz}$, which would make $S_{\phi\phi}$, $S_{JJ}$, and $\kappa$ frequency-dependent.  This $1/f$ noise can be avoided by up-converting the signal, using for instance a superconducting capacitor bridge transducer \cite{Cinquegrana1993, Paik2016} or a superconducting inductance bridge transducer \cite{Paik1986}.  In our subsequent estimates, we will assume that this upconversion can be achieved, so that we can treat $\kappa$, $S_{\phi\phi}$ and $S_{JJ}$ as frequency-independent.  In this case, the combination of all noise sources can be written as
\begin{align}
    S_{BB}^\mathrm{tot}&=S_{BB}^\mathrm{th}+S_{BB}^\mathrm{imp}+S_{BB}^\mathrm{back}\nl
    =\frac{2\rho\left(4m\gamma T+\kappa\tilde\eta^{-2}|\chi(\omega)|^{-2}+\kappa\tilde\eta^2\right)}{3m^2\omega_0^2},
    \label{eq:total_noise}
\end{align}
where $\tilde\eta=\eta\sqrt[4]{\frac{S_{JJ}}{S_{\phi\phi}}}$.

We can vary the relative sizes of these contributions by changing the coupling $\eta$.  As mentioned above, there is, however, a trade-off between imprecision noise and back-action noise, when we do so.  Since both back-action and thermal noise are frequency-independent, there is no benefit in decreasing $\eta$ beyond the point where thermal noise dominates over back-action noise.  Thus, (if possible) we should always take
\begin{equation}
    \tilde\eta\geq\sqrt{\frac{4m\gamma T}\kappa}.
    \label{eq:res_eta}
\end{equation}

\begin{figure}[t]
\includegraphics[width=0.99\columnwidth]{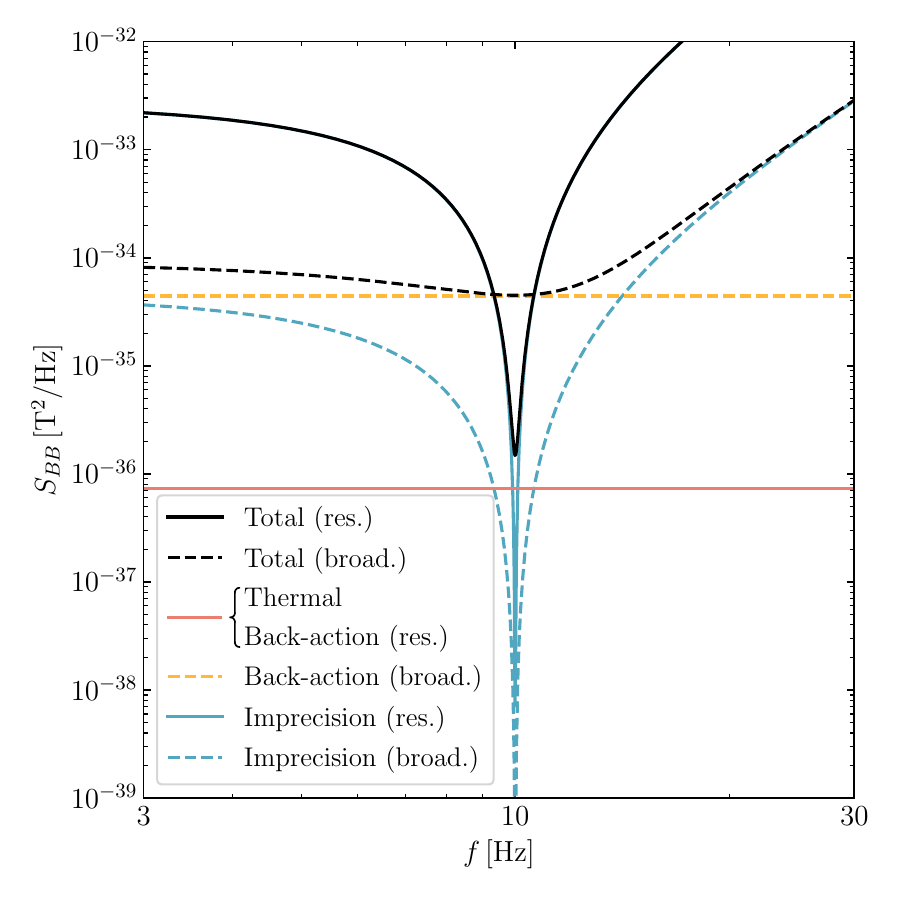}
\caption{\label{fig:noise}%
    Noise curves for resonant (solid) and broadband (dashed) choices of $\eta$.  The black curves show the total noise, while the colored curves show the thermal (red), back-action (orange), and imprecision (blue) noise contributions.  To compute these curves, we use the same parameter values as in \eqref{thermal_noise_est}, except with $f_0=10\,\mathrm{Hz}$ and $\kappa=5$.  Note that in the case of the resonant choice, the back-action noise coincides with the thermal noise.  Moreover, thermal noise is independent of the choice of $\eta$.  Therefore, the red curve represents the thermal noise in both cases, as well as the back-action noise in the resonant case.}
\end{figure}

On the other hand, at low frequencies $\omega\ll\omega_0$, we have $\chi\approx1/(m\omega_0^2)$, and so imprecision noise (for a fixed test mass $m$) is frequency-independent as well.  Therefore, increasing $\eta$ beyond the point where back-action noise dominates over imprecision noise is not beneficial at frequencies lower than the trapping frequency.%
\footnote{Increasing $\eta$ further can still be beneficial at high frequencies $\omega\gg\omega_0$, but the sensitivity in this regime degrades rapidly (see \figref{noise}), so we do not consider increasing $\eta$ further to be a productive way of improving sensitivity.  Instead, if one wants to probe higher frequencies, it is better to increase $\omega_0$.}
In other words, we also want
\begin{equation}
    \tilde\eta\leq\frac1{\sqrt{|\chi(0)|}}=\sqrt m\omega_0.
    \label{eq:broad_eta}
\end{equation}
We expect that it should generally be possible to saturate this upper bound by appropriate design of the readout; e.g., see the supplemental material of \citeR{Hofer2023}.  Meanwhile, the coupling can always be decreased by worsening the readout efficiency.  Therefore, we expect that $\tilde\eta$ can be varied across the entire range from \eqref{res_eta} to \eqref{broad_eta}.

In our sensitivity calculations below, we consider two choices of $\eta$, corresponding to the limiting cases in \eqref{res_eta} and (\ref{eq:broad_eta}).  We refer to these as the ``resonant" and ``broadband" choices, respectively, as the former maximizes sensitivity at $\omega=\omega_0$,%
\footnote{The resonant sensing scheme takes advantage of the low imprecision noise around the resonance frequency $\omega_0$ within a narrow frequency range of $\sim\max(\gamma, 1/t_\mathrm{int})$ [the latter describes Fourier broadening]. This requires that resonance frequency drifts within the integration time are small compared with $\max(\gamma, 1/t_\mathrm{int})$. For instance, in \citeR{Hofer2023}, the current in the trap coils was unstable, causing $\omega_0$ to drift and preventing the on-resonance sensing enhancement to be fully demonstrated. Such drifts can be mitigated by using persistent superconducting currents \cite{goodkind} in the trap coils.}
while the latter maximizes sensitivity at $\omega\ll\omega_0$.  \figref{noise} shows sample noise curves for these two different choices, along with the individual noise contributions in each case.

\subsection{Projections}

\begin{figure*}[t]
\includegraphics[width=0.49\textwidth]{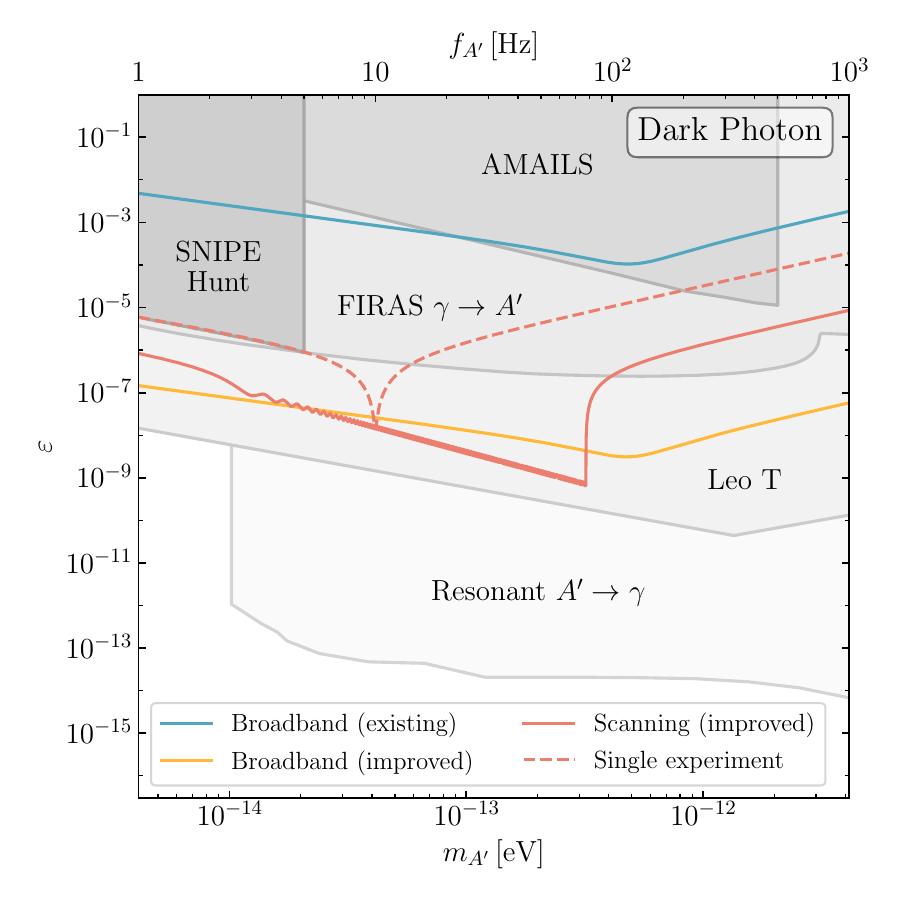}
\includegraphics[width=0.49\textwidth]{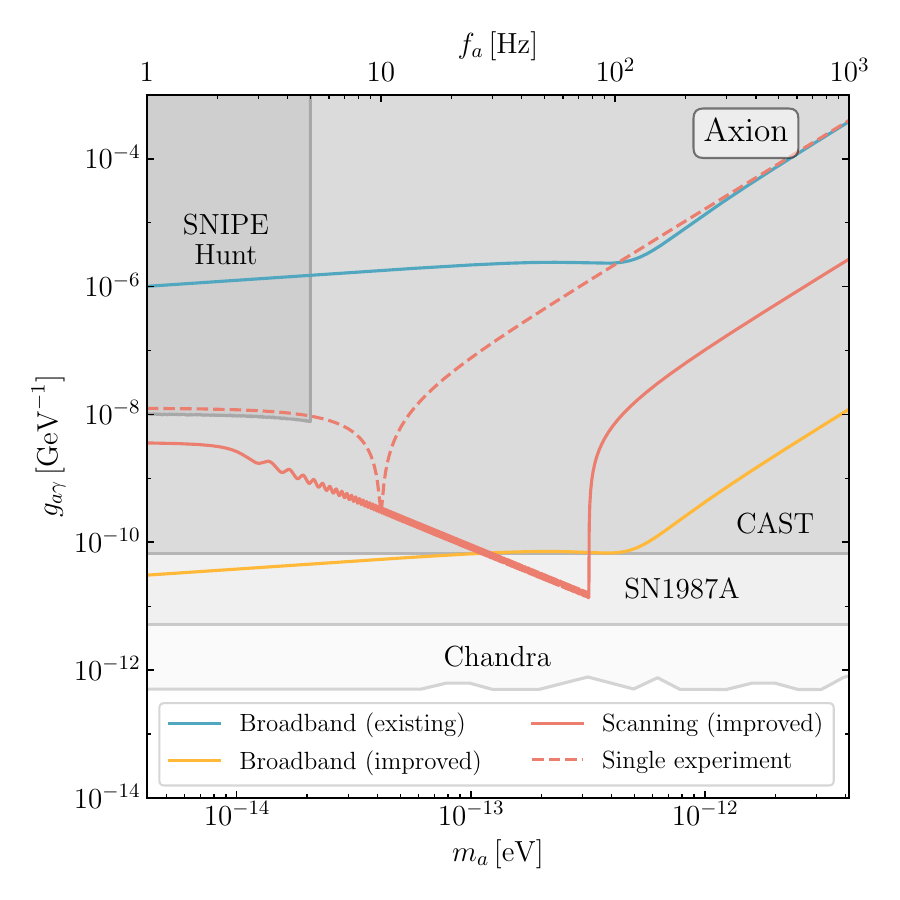}
\caption{\label{fig:sensitivity}%
    Sensitivity of levitated superconductors to DPDM (left) and axion DM (right).  The blue curves show the sensitivity achievable with parameters representative of an existing setup (with increased readout efficiency), as in \citeR{Hofer2023}. In contrast, the orange and red curves show the sensitivity of a new setup with improved parameters, including a larger hollow SCP.  The parameter values for both setups are shown in Table~\ref{tab:parameters}.  The blue and orange curves consider a single experiment conducted for $t_\mathrm{int}=1\,\mathrm{yr}$, using a trapping frequency of $f_0=100\,\mathrm{Hz}$ and the ``broadband" choice of coupling $\eta$ (see main text).  The dashed red curves represent a single experiment conducted for $t_\mathrm{int}=t_\mathrm{coh}\sim30\,\mathrm{hr}$, using a trapping frequency of $f_0=10\,\mathrm{Hz}$ and the ``resonant" choice of $\eta$.  The solid red curves show the aggregate sensitivity of scanning this resonant setup over many trapping frequencies from $f_0=3\,\mathrm{Hz}$ to $77\,\mathrm{Hz}$ (so that the total integration time is $1\,\mathrm{yr}$).  We also show existing constraints in various shades of grey (see main text for descriptions).  Laboratory constraints (SNIPE Hunt, AMAILS, and CAST) are shown in darker shades of grey, while astrophysical/cosmological constraints are shown in lighter shades.  These sensitivity curves demonstrate that existing levitation setups with improved readout efficiencies are comparable to other laboratory probes of DPDM. In addition, a focused, dedicated setup can achieve the leading sensitivity amongst such probes of both DPDM and axion DM.
    }
\end{figure*}

With these choices for $\eta$, we can project sensitivity curves for DPDM and axion DM using our proposed setups.  The simpler case is to utilize the broadband choice.  In this case, good sensitivity to a wide range of masses can be achieved by running a single experiment with a fixed resonant frequency $\omega_0$.  From \eqref{total_noise} with the choice of $\eta$ as in \eqref{broad_eta}, we can see that, in the regime where imprecision and backaction noise dominate over thermal noise (see \figref{noise}), the total noise at low frequencies $\omega\ll\omega_0$ is independent of $\omega_0$.  Therefore, our choice of the resonant frequency will not affect our sensitivity at low frequencies (in the DPDM case),%
\footnote{In the axion case, the signal also scales with $\omega_0$ [see \eqref{axion_signal}].  Therefore in the axion case, larger $\omega_0$ will actually improve our sensitivity at low frequencies.  We do note, however, that as per \eqref{Bmax}, increasing $\omega_0$ also increases the maximum magnetic field on the surface of the SCP, $B_\mathrm{max}$.  This means that larger $\omega_0$ more strongly constrains the size of the SCP.}
and so it is best to choose $\omega_0$ as large as possible to minimize the frequency range that suffers the high-frequency suppression.  Our projections in \figref{sensitivity} take $f_0=100\,\mathrm{Hz}$.

For short integration times $t_\mathrm{int}$, the signal-to-noise ratio (SNR) for such an experiment can be determined as
\begin{equation}
    \mathrm{SNR}=\frac{B_\mathrm{DM}^2}{2S_{BB}^\mathrm{tot}/t_\mathrm{int}},
    \label{eq:SNR_coh}
\end{equation}
where $B_\mathrm{DM}$ is the magnetic field signal in \eqref[s]{DPDM_signal} and (\ref{eq:axion_signal}) for the DPDM case and the axion case, respectively.
However, \eqref[s]{DPDM_wave} and (\ref{eq:axion_wave}) [and so also \eqref[s]{DPDM_signal} and (\ref{eq:axion_signal})] are only valid on timescales $t_\mathrm{int}$ shorter than the coherence time $t_\mathrm{coh}\sim2\pi/(m_\mathrm{DM}v_\mathrm{DM}^2)\sim10^6/f_\mathrm{DM}$ of the DM. On timescales longer than this, the amplitudes $A'_0$ and $a_0$ in \eqref[s]{DPDM_signal} and (\ref{eq:axion_signal}) vary stochastically (see footnote~\ref{ftnt:norm} for a discussion of their normalization).  For $t_\mathrm{int}>t_\mathrm{coh}$, we can then treat each coherence time as an independent experiment.  To get the SNR for the full $t_\mathrm{int}$, we sum the SNRs from each individual coherence time in quadrature~\cite{Budker:2013hfa}
\begin{equation}
    \mathrm{SNR}=\frac{B_\mathrm{DM}^2}{2S_{BB}^\mathrm{tot}/t_\mathrm{coh}}\cdot\sqrt{\frac{t_\mathrm{int}}{t_\mathrm{coh}}}.
    \label{eq:SNR}
\end{equation}
The blue and orange curves labeled ``broadband" in \figref{sensitivity} show the projected sensitivities to DPDM and axion DM, respectively.  These are computed by setting $\mathrm{SNR}=3$ in \eqref{SNR}, utilizing the broadband choice for $\eta$ in $S_{BB}^\mathrm{tot}$, fixing a trapping frequency $f_0=100\,\mathrm{Hz}$, and taking an integration time of $t_\mathrm{int}=1\,\mathrm{yr}$.

The blue curves in \figref{sensitivity} show the sensitivities which can be achieved with parameters representative of an existing levitation setup, such as in \citeR{Hofer2023}, if they can improve their coupling strength close to the bound in \eqref{broad_eta}.  In principle, the only other modification required for such a setup to be sensitive to ultralight DM is to shift the trap off-center within the shield.  The orange curves show the sensitivities that can be achieved with an improved setup.  Most notably, this setup considers a SCP that is much larger and hollow, along with a reduced dissipation rate and larger apparatus dimensions.  The parameter values used for these setups are shown in Table~\ref{tab:parameters}.  In the DPDM case, we consider only the sensitivity to the $z$-component of $\bm A'$, for simplicity, but we note that marginally better sensitivity could be achieved by considering all three components.  In the axion case, we take the trap to be located at a position $\bm r_0=(0.7L,0.8L,0.5L)$ within the shield.

\begin{table}
    \centering
    \begin{tabular}{c c c}
        \hline\hline
        Parameter & Existing & Improved\\ 
        \hline
        SCP mass $m$&$10\,\mu\mathrm g$&$1\,\mathrm g$\\
        SCP density $\rho$&$10\,\mathrm{g/cm}^3$&$0.1\,\mathrm{g/cm}^3$\\
        Dissipation rate $\gamma$&~~$2\pi\cdot10^{-5}\,\mathrm{Hz}$~~&~~$2\pi\cdot10^{-8}\,\mathrm{Hz}$~~\\
        Temperature $T$&\multicolumn{2}{c}{$10\,\mathrm{mK}$}\\
        SQUID energy resolution $\kappa$&\multicolumn{2}{c}{5}\\
        \hline
        Distance from axis $r$&$10\,\mathrm{cm}$&$1\,\mathrm{m}$\\
        \hline
        Shield dimension $L$&$10\,\mathrm{cm}$&$1\,\mathrm{m}$\\
        Coil radius $R$&$1\,\mathrm{cm}$&$10\,\mathrm{cm}$\\
        Coil separation $h$&$1\,\mathrm{cm}$&$10\,\mathrm{cm}$\\
        \hline\hline
    \end{tabular}
    \caption{\label{tab:parameters}%
    Parameters used to compute the sensitivity curves in \figref{sensitivity}.  One column shows parameter values representative of an existing setup, as in \citeR{Hofer2023}, while the other shows parameter values for an improved setup.  The first set of parameters is common to both the DPDM and axion DM scenarios.  The parameter $r$ is relevant in the DPDM scenario [as in \eqref{DPDM_signal}], while the parameters $L$, $R$, and $h$ are relevant in the axion DM scenario [as in \eqref{axion_signal}].
    }
\end{table}

In the resonant case, we achieve excellent sensitivity near $\omega_0$, but worse sensitivity away from it.  We will, therefore, need to perform several experiments of shorter durations, each with a different $\omega_0$.  The trapping frequency can be scanned e.g., by varying the current running through the coils, which will change $b_0$.  Each such experiment will only effectively probe some small range $\delta\omega$ of frequency space.  We can estimate this width by determining when $S_{BB}^\mathrm{tot}$ doubles in size, that is
\begin{equation}
    S_{BB}^\mathrm{tot}\left(\omega_0+\frac{\delta\omega}2\right)=2S_{BB}^\mathrm{tot}(\omega_0)
\end{equation}
(using the resonant choice \eqref{res_eta} for $\eta$).  Assuming thermal noise and backaction noise dominate over imprecision noise at $\omega=\omega_0$ (see \figref{noise}), this implies
\begin{align}
    \label{eq:width}
    \delta\omega&=\frac{4\sqrt2\gamma T}{\kappa\omega_0}\\
    &\sim2\pi\cdot0.2\,\mathrm{Hz}\left(\frac\gamma{2\pi\cdot10^{-8}\,\mathrm{Hz}}\right)\nl\cdot\left(\frac T{10\,\mathrm{mK}}\right)\left(\frac5\kappa\right)\left(\frac{10\,\mathrm{Hz}}{f_0}\right).
\end{align}
We will, therefore, need to run experiments at several trapping frequencies $\omega_i$, separated from each other by roughly $\delta\omega_i=\delta\omega(\omega_0=\omega_i)$ [so the $\omega_i$ values will be closer together at higher frequencies].  As our sensitivity improves more slowly for $t_\mathrm{int}>t_\mathrm{coh}$ [see \eqref{SNR}], we will fix the integration time of each experiment to be $t_{\mathrm{int},i}=t_\mathrm{coh}(m_\mathrm{DM}=\omega_i)$.%
\footnote{Note that there is also a lower bound on $t_{\mathrm{int},i}$.  This is because an applied AC force takes time to ring up the position oscillations of the SCP fully.  In other words, the imprecision noise $S_{BB}^\mathrm{imp}$ in \eqref{measurement_noise} only receives the $|\chi(0)/\chi(\omega_0)|^{-2}=(\gamma/\omega_0)^2$ suppression on resonance for times $t_\mathrm{int}>\frac{2\pi}\gamma$.  Since the system rings up linearly at short times, this suppression should instead be $(2\pi/\omega_0t_\mathrm{int})^2$ for shorter times.  In order for imprecision noise to be subdominant to thermal and back-action noise on resonance, we must have
\begin{align}
    t_{\mathrm{int},i}>\frac{\pi\kappa\omega_0}{2\sqrt2\gamma T}&\sim40\,\mathrm{s}\left(\frac\kappa5\right)\left(\frac{f_0}{100\,\mathrm{Hz}}\right)\nl
    \cdot\left(\frac{2\pi\cdot10^{-8}\,\mathrm{Hz}}\gamma\right)\left(\frac {10\,\mathrm{mK}}T\right).
\end{align}
By comparison, $t_\mathrm{coh}\sim10^4\,\mathrm{s}$ for $m_\mathrm{DM}=2\pi\cdot100\,\mathrm{Hz}$, so this bound is satisfied for the entire projected sensitivity curve appearing in \figref{sensitivity}.  Moreover, the sensitivity width in \eqref{width} is unaffected because $\delta\omega\gg2\pi/t_\mathrm{coh}$, so the broadening of the signal due to finite integration time is negligible.}
If we wish to scan over a total frequency range of $\Delta\omega$, the total integration time will then be
\begin{align}
    \sum_it_{\mathrm{int},i}&=\sum_i\frac{\kappa\pi}{2\sqrt2\gamma Tv_\mathrm{DM}^2}\delta\omega_i\nl
    =\frac{\kappa\pi}{2\sqrt2\gamma Tv_\mathrm{DM}^2}\Delta\omega\\
    &\sim1\,\mathrm{yr}\left(\frac\kappa 5\right)\left(\frac{2\pi\cdot10^{-8}\,\mathrm{Hz}}\gamma\right)\nl\cdot\left(\frac {10\,\mathrm{mK}}T\right)\left(\frac{\Delta f}{74\,\mathrm{Hz}}\right).
\end{align}

The solid red curves in \figref{sensitivity} show the projected sensitivities for this scanning scheme (and the ``improved" parameters mentioned above).  We scan from $f_0=3\,\mathrm{Hz}$ up to $77\,\mathrm{Hz}$, so that the total integration time is 1\,yr.  The SNR for the experiment with trapping frequency $\omega_i$ is calculated using \eqref{SNR}, with $t_\mathrm{int}=t_\mathrm{coh}(\omega_i)$ and the resonant choice for $\eta$ in $S_{BB}^\mathrm{tot}$.  The SNRs of the individual experiments are then combined in quadrature,\footnote{Adding the SNRs in quadrature is necessary when the DM signal is not coherent from one experiment to the next.  Because the experiment with trapping frequency $f_i$ integrates for $t_\mathrm{coh}(\omega_i)$, the SNRs must be summed in quadrature for $m_\mathrm{DM}\geq\omega_i$.  In principle, the SNRs can be summed linearly for $m_\mathrm{DM}<\omega_i$, but we expect the gain to be marginal as the sensitivity for masses $m_\mathrm{DM}\geq2\pi\cdot3\,\mathrm{Hz}$ is dominated by the single experiment with trapping frequency near $m_\mathrm{DM}$.  For simplicity, we therefore sum in quadrature for all masses.} i.e.
\begin{equation}
    \mathrm{SNR}^2=\sum_i\mathrm{SNR}_i^2,
\end{equation}
where the index $i$ runs over the individual experiments.  The sensitivity in \figref{sensitivity} takes a total $\mathrm{SNR}=3$.  For $3\,\mathrm{Hz}<f_\mathrm{DM}<77\,\mathrm{Hz}$, the sensitivity is dominated by the peak sensitivity of the experiment with trapping frequency $f_0=f_\mathrm{DM}$.  Outside this frequency range, the low/high-frequency tails of several experiments contribute to the combined sensitivity.  The dashed red curves also show the sensitivities of a single experiment with $f_0=10\,\mathrm{Hz}$.

In \figref{sensitivity}, we also show existing constraints in various shades of grey.%
\footnote{Several of these limits were acquired from \citeR[s]{Oharegithub,Caputo:2021eaa}.  See also \citeR[s]{Witte:2020rvb,Caputo:2020bdy,escudero2023axion,Cardoso_2018} for other limits in this mass range which are not shown here, and \citeR{bloch2023curl} for a brief discussion of the caveats regarding those limits.}
The DPDM constraints include limits from: unshielded magnetometer measurements by the SNIPE Hunt collaboration~\cite{sulai2023hunt}; magnetometer measurements taken inside a shielded room by the AMAILS collaboration~\cite{jiang2023search}; non-observation of CMB-photon conversion into (non-DM) dark photons by the FIRAS instrument~\cite{Caputo:2020bdy}; heating of the dwarf galaxy Leo T~\cite{Wadekar:2019xnf}; and resonant conversion of DPDM during the dark ages~\cite{McDermott:2019lch}.  The axion constraints include limits from: SNIPE Hunt; the CAST helioscope search for solar axions~\cite{Anastassopoulos:2017ftl}; non-observation of gamma rays in coincidence with SN1987A~\cite{Hoof_2023}; and X-ray observations of the quasar H1821+643 from the Chandra telescope~\cite{10.1093/mnras/stab3464}.  Laboratory constraints (SNIPE Hunt, AMAILS, and CAST) are shown in darker shades of grey, while astrophysical/cosmological constraints are shown in lighter shades.

\section{Discussion}
\label{sec:discussion}

In this work, we explored the prospect of utilizing magnetically levitated superconductors to search for ultralight DM at frequencies below kHz.  If ultralight DM couples to electromagnetism, it can source an oscillating magnetic field inside an experimental apparatus.  Various experimental methods exist to probe such magnetic field signals at high frequencies, but few existing or proposed experiments are sensitive to DM with masses corresponding to frequencies $f_\mathrm{DM}\lesssim\mathrm{kHz}$.  We showed that levitated superconductors can function as excellent magnetometers, which are sensitive to signals in the Hz to kHz frequency range.  This makes them well suited to detect ultralight DM in the $4\times10^{-15}\lesssim m_\mathrm{DM}\lesssim4\times10^{-12}\,\mathrm{eV}$ mass range.

A superconductor immersed in a magnetic field configuration will tend to settle at the point of lowest magnetic field.  This fact can trap a SCP at the center of a quadrupole magnetic field.  Ultralight DM can source a nonzero oscillating magnetic field signal near the trap if such a trap is located off-center within a magnetic shield.  This DM-sourced field can perturb the equilibrium position of the SCP, leading to oscillatory motion.  If the frequency of this magnetic field signal matches the trapping frequency (typically in the Hz to kHz range), then the motion can be resonantly enhanced.  This makes levitated superconductors unique among axion and dark-photon experiments in that they can resonantly search for these DM candidates for masses $m_\mathrm{DM}\lesssim10^{-12}\,\mathrm{eV}$.

We discussed three primary noise sources for a levitated SCP experiment: thermal noise, imprecision noise, and back-action noise.  The first is fixed by the experiment's dissipation rate and temperature, while the parameters of the readout system fix the latter two.  In particular, a trade-off exists between imprecision and back-action noise, which allows for two different operation schemes of a levitated SCP experiment.  In the ``broadband" scheme (the blue and orange curves in \figref{sensitivity}), sensitivity to a wide range of frequencies is maximized by equating back-action noise with below-resonance imprecision noise.  In this case, a single experiment run for a long duration can achieve excellent sensitivity at many DM masses.  In the ``resonant" scheme (the red curves in \figref{sensitivity}), sensitivity on-resonance is maximized by equating thermal and back-action noise.  In this case, several shorter-duration experiments are required to scan a large range of DM masses.

\figref{sensitivity} shows that, with a strongly coupled readout, existing levitation experiments (blue curves) can already achieve sensitivity to DPDM comparable to other laboratory experiments in this mass range.  A dedicated setup (orange and red curves) using larger hollow spheres, a lower dissipation rate, and a larger apparatus can achieve even better sensitivity.  In particular, in the DPDM case, it can exceed the existing laboratory constraints and approach the best astrophysical heating constraints, while in the axion DM case, it can be the best laboratory probe and approach constraints from SN1987A.  Since these astrophysical constraints can depend quite sensitively on the modeling of complex systems, it is valuable to have complementary laboratory probes.  Both the broadband (orange) and scanning (red) schemes enable good sensitivities for this improved setup.

While our projections already show that levitated superconductors can be promising ultralight DM detectors, this technique could potentially be improved in several ways.  Firstly, the thermal noise floor can be decreased by further lowering the temperature of the system.  Doing so will likely not affect the sensitivity of our broadband scheme, as thermal noise is typically subdominant, but it can improve the sensitivity of the resonant scheme.  We also note that utilizing an array of sensors and/or a squeezed readout can further improve the sensitivity and scan rate of the experiment, which demands further investigation~\cite{Brady:2022bus,Brady:2022qne,Xia:2022rql}.

Secondly, different geometries of the levitation apparatus can be considered.  Our projections in \figref{sensitivity} assume a spherical SCP levitated between circular anti-Helmholtz-like coils, which would result in $b_{xx}=b_{yy}=-\frac12b_{zz}$.  By utilizing elliptic coils, the degeneracy between $b_{xx}$ and $b_{yy}$ can be broken, potentially allowing for frequency hierarchies $b_{xx}\ll b_{zz}$.  This would enable the apparatus to probe lower frequencies while maintaining a small displacement of the equilibrium position due to gravity [see \eqref{grav_displacement}].  A coaxial levitation geometry could also enable a hierarchy between the radial and axial frequencies~\cite{Griggs2017}.  Additionally, the SCP shape can be varied to decrease the effective density even further below the densities used in the improved setup of \figref{sensitivity}.  For instance, a SCP in the shape of a ring can have a much smaller mass than a sphere of the same effective volume~\cite{Navau2021,GutierrezLatorre2020}.

Finally, a larger signal can be created in the axion case by utilizing a larger static magnetic field for axion-photon conversion.  In this work, we have assumed that the magnetic field allowing the axion DM to convert is the same magnetic field that traps the SCP.  However, an additional magnetic field can be applied, which enhances the axion-photon conversion rate without affecting the trapping physics.  The calculation in \appref{axion} shows that the axion signal in the vicinity of the trap is affected by static magnetic fields sourced from anywhere within the shield (not just the magnetic field sourced by the trap).  It is, therefore, plausible that a large static magnetic field can be sourced at the opposite end of the shield so that it does not significantly affect the operation of the levitated SCP apparatus, but it has a large effect on the axion magnetic field signal.  We leave a detailed study of this idea to future work.

\begin{acknowledgments}
We thank Asher Berlin, Dan Carney, Yifan Chen, Roni Harnik, Junwu Huang, Rafael Lang, Jacob Taylor, and Yue Zhao for their helpful discussions.
Z.L. and S.K. are supported in part by the U.S. Department of Energy, Office of Science, National Quantum Information Science Research
Centers, Superconducting Quantum Materials and Systems Center (SQMS) under contract number DE-AC02-07CH11359. Z.L. is supported in part the DOE grant DE-SC0011842.
Z.L. and S.K. acknowledge the support of the Aspen Center for Physics, supported by National Science Foundation grant PHY-2210452, where part of this work was completed.
Z.L. and S.K. acknowledge support from the Simons Foundation Targeted Grant 920184 to the Fine Theoretical Physics Institute, which positively impacted our work.
G.H.\ acknowledges support from the Swedish Research Council (Grant No.\ 2020-00381).  The code used for this research is made publicly available through Github~\cite{github} under CC-BY-NC-SA.
\end{acknowledgments}

\bibliographystyle{JHEP}
\bibliography{references.bib}

\providecommand{\href}[2]{#2}\begingroup\justify\begin{thebibliography}{100}

\bibitem{Arias:2012az}
P.~Arias, D.~Cadamuro, M.~Goodsell, J.~Jaeckel, J.~Redondo and A.~Ringwald,
  \emph{{WISPy Cold Dark Matter}},
  \href{https://dx.doi.org/10.1088/1475-7516/2012/06/013}{\emph{JCAP} {\bf 06}
  (2012) 013} [\href{https://arxiv.org/abs/1201.5902}{{\tt arXiv:1201.5902}}].

\bibitem{kimball2022search}
D.~F. Jackson~Kimball and K.~van Bibber, \emph{{The Search for Ultralight
  Bosonic Dark Matter}}.
\newblock Springer, 2022,
  \href{https://dx.doi.org/10.1007/978-3-030-95852-7}{10.1007/978-3-030-95852-7}.

\bibitem{Evans:2018bqy}
N.~W. Evans, C.~A.~J. O'Hare and C.~McCabe, \emph{{Refinement of the standard
  halo model for dark matter searches in light of the Gaia Sausage}},
  \href{https://dx.doi.org/10.1103/PhysRevD.99.023012}{\emph{Phys. Rev. D} {\bf
  99} (2019) 023012} [\href{https://arxiv.org/abs/1810.11468}{{\tt
  arXiv:1810.11468}}].

\bibitem{lin2018self}
S.-C. Lin, H.-Y. Schive, S.-K. Wong and T.~Chiueh, \emph{Self-consistent
  construction of virialized wave dark matter halos},
  \href{https://dx.doi.org/10.1103/PhysRevD.97.103523}{\emph{Phys. Rev. D} {\bf
  97} (2018) 103523}.

\bibitem{Centers:2019dyn}
G.~P. Centers, J.~W. Blanchard, J.~Conrad, N.~L. Figueroa, A.~Garcon, A.~V.
  Gramolin et~al., \emph{Stochastic fluctuations of bosonic dark matter},
  \href{https://dx.doi.org/10.1038/s41467-021-27632-7}{\emph{Nat. Comm.} {\bf
  12} (2021) 7321}.

\bibitem{Preskill:1982cy}
J.~Preskill, M.~B. Wise and F.~Wilczek, \emph{{Cosmology of the Invisible
  Axion}}, \href{https://dx.doi.org/10.1016/0370-2693(83)90637-8}{\emph{Phys.
  Lett. B} {\bf 120} (1983) 127--132}.

\bibitem{Abbott:1982af}
L.~Abbott and P.~Sikivie, \emph{{A Cosmological Bound on the Invisible Axion}},
  \href{https://dx.doi.org/10.1016/0370-2693(83)90638-X}{\emph{Phys. Lett. B}
  {\bf 120} (1983) 133--136}.

\bibitem{Dine:1982ah}
M.~Dine and W.~Fischler, \emph{{The Not So Harmless Axion}},
  \href{https://dx.doi.org/10.1016/0370-2693(83)90639-1}{\emph{Phys. Lett. B}
  {\bf 120} (1983) 137--141}.

\bibitem{Gra15}
P.~W. Graham, D.~E. Kaplan and S.~Rajendran, \emph{Cosmological relaxation of
  the electroweak scale},
  \href{https://dx.doi.org/10.1103/PhysRevLett.115.221801}{\emph{Phys. Rev.
  Lett.} {\bf 115} (2015) 221801} [\href{https://arxiv.org/abs/1504.07551}{{\tt
  arXiv:1504.07551}}].

\bibitem{co2020predictions}
R.~T. Co, L.~J. Hall and K.~Harigaya, \emph{Predictions for axion couplings
  from {ALP} cogenesis},
  \href{https://dx.doi.org/10.1007/JHEP01(2021)172}{\emph{JHEP} {\bf 01} (2021)
  172} [\href{https://arxiv.org/abs/2006.04809}{{\tt arXiv:2006.04809}}].

\bibitem{Holdom:1986ag}
B.~Holdom, \emph{{Two $U(1)$'s and $\epsilon$ Charge Shifts}},
  \href{https://dx.doi.org/10.1016/0370-2693(86)91377-8}{\emph{Phys. Lett. B}
  {\bf 166} (1986) 196--198}.

\bibitem{Nelson:2011sf}
A.~E. Nelson and J.~Scholtz, \emph{{Dark Light, Dark Matter and the
  Misalignment Mechanism}},
  \href{https://dx.doi.org/10.1103/PhysRevD.84.103501}{\emph{Phys. Rev. D} {\bf
  84} (2011) 103501} [\href{https://arxiv.org/abs/1105.2812}{{\tt
  arXiv:1105.2812}}].

\bibitem{Graham:2015rva}
P.~W. Graham, J.~Mardon and S.~Rajendran, \emph{{Vector Dark Matter from
  Inflationary Fluctuations}},
  \href{https://dx.doi.org/10.1103/PhysRevD.93.103520}{\emph{Phys. Rev. D} {\bf
  93} (2016) 103520} [\href{https://arxiv.org/abs/1504.02102}{{\tt
  arXiv:1504.02102}}].

\bibitem{Peccei:1977hh}
R.~Peccei and H.~R. Quinn, \emph{{CP Conservation in the Presence of
  Instantons}},
  \href{https://dx.doi.org/10.1103/PhysRevLett.38.1440}{\emph{Phys. Rev. Lett.}
  {\bf 38} (1977) 1440--1443}.

\bibitem{Weinberg:1977ma}
S.~Weinberg, \emph{{A New Light Boson?}},
  \href{https://dx.doi.org/10.1103/PhysRevLett.40.223}{\emph{Phys. Rev. Lett.}
  {\bf 40} (1978) 223--226}.

\bibitem{Wilczek:1977pj}
F.~Wilczek, \emph{{Problem of Strong $P$ and $T$ Invariance in the Presence of
  Instantons}},
  \href{https://dx.doi.org/10.1103/PhysRevLett.40.279}{\emph{Phys. Rev. Lett.}
  {\bf 40} (1978) 279--282}.

\bibitem{cvetivc1996implications}
M.~Cveti{\v{c}} and P.~Langacker, \emph{{Implications of Abelian extended gauge
  structures from string models}},
  \href{https://dx.doi.org/10.1103/PhysRevD.54.3570}{\emph{Phys. Rev. D} {\bf
  54} (1996) 3570} [\href{https://arxiv.org/abs/hep-ph/9511378}{{\tt
  arXiv:hep-ph/9511378}}].

\bibitem{Svrcek:2006yi}
P.~Svrcek and E.~Witten, \emph{{Axions In String Theory}},
  \href{https://dx.doi.org/10.1088/1126-6708/2006/06/051}{\emph{JHEP} {\bf 06}
  (2006) 051} [\href{https://arxiv.org/abs/hep-th/0605206}{{\tt
  arXiv:hep-th/0605206}}].

\bibitem{Arvanitaki:2009fg}
A.~Arvanitaki, S.~Dimopoulos, S.~Dubovsky, N.~Kaloper and J.~March-Russell,
  \emph{{String Axiverse}},
  \href{https://dx.doi.org/10.1103/PhysRevD.81.123530}{\emph{Phys. Rev. D} {\bf
  81} (2010) 123530} [\href{https://arxiv.org/abs/0905.4720}{{\tt
  arXiv:0905.4720}}].

\bibitem{Sikivie:1983ip}
P.~Sikivie, \emph{{Experimental Tests of the Invisible Axion}},
  \href{https://dx.doi.org/10.1103/PhysRevLett.51.1415}{\emph{Phys. Rev. Lett.}
  {\bf 51} (1983) 1415--1417}.

\bibitem{ehret2010new}
K.~Ehret, M.~Frede, S.~Ghazaryan, M.~Hildebrandt, E.-A. Knabbe, D.~Kracht
  et~al., \emph{{New ALPS results on hidden-sector lightweights}},
  \href{https://dx.doi.org/10.1016/j.physletb.2010.04.066}{\emph{Phys. Lett. B}
  {\bf 689} (2010) 149--155} [\href{https://arxiv.org/abs/1004.1313}{{\tt
  arXiv:1004.1313}}].

\bibitem{Wagner:2010mi}
{\scshape ADMX Collaboration}, A.~Wagner, G.~Rybka, M.~Hotz, L.~J. Rosenberg,
  S.~J. Asztalos, G.~Carosi et~al., \emph{{A Search for Hidden Sector Photons
  with ADMX}},
  \href{https://dx.doi.org/10.1103/PhysRevLett.105.171801}{\emph{Phys. Rev.
  Lett.} {\bf 105} (2010) 171801} [\href{https://arxiv.org/abs/1007.3766}{{\tt
  arXiv:1007.3766}}].

\bibitem{Redondo:2010dp}
J.~Redondo and A.~Ringwald, \emph{{Light shining through walls}},
  \href{https://dx.doi.org/10.1080/00107514.2011.563516}{\emph{Contemp. Phys.}
  {\bf 52} (2011) 211--236} [\href{https://arxiv.org/abs/1011.3741}{{\tt
  arXiv:1011.3741}}].

\bibitem{Horns:2012jf}
D.~Horns, J.~Jaeckel, A.~Lindner, A.~Lobanov, J.~Redondo and A.~Ringwald,
  \emph{{Searching for WISPy Cold Dark Matter with a Dish Antenna}},
  \href{https://dx.doi.org/10.1088/1475-7516/2013/04/016}{\emph{JCAP} {\bf 04}
  (2013) 016} [\href{https://arxiv.org/abs/1212.2970}{{\tt arXiv:1212.2970}}].

\bibitem{Betz_2013}
M.~Betz, F.~Caspers, M.~Gasior, M.~Thumm and S.~W. Rieger, \emph{First results
  of the {CERN} resonant weakly interacting sub-{eV} particle search
  ({CROWS})},
  \href{https://dx.doi.org/10.1103/physrevd.88.075014}{\emph{Physical Review D}
  {\bf 88} (Oct 2013) }.

\bibitem{Graham:2014sha}
P.~W. Graham, J.~Mardon, S.~Rajendran and Y.~Zhao, \emph{{Parametrically
  enhanced hidden photon search}},
  \href{https://dx.doi.org/10.1103/PhysRevD.90.075017}{\emph{Phys. Rev. D} {\bf
  90} (2014) 075017} [\href{https://arxiv.org/abs/1407.4806}{{\tt
  arXiv:1407.4806}}].

\bibitem{Chaudhuri:2014dla}
S.~Chaudhuri, P.~W. Graham, K.~Irwin, J.~Mardon, S.~Rajendran and Y.~Zhao,
  \emph{{Radio for hidden-photon dark matter detection}},
  \href{https://dx.doi.org/10.1103/PhysRevD.92.075012}{\emph{Phys. Rev. D} {\bf
  92} (2015) 075012} [\href{https://arxiv.org/abs/1411.7382}{{\tt
  arXiv:1411.7382}}].

\bibitem{graham2016dark}
P.~W. Graham, D.~E. Kaplan, J.~Mardon, S.~Rajendran and W.~A. Terrano,
  \emph{Dark matter direct detection with accelerometers},
  \href{https://dx.doi.org/10.1103/PhysRevD.93.075029}{\emph{Phys. Rev. D} {\bf
  93} (2016) 075029} [\href{https://arxiv.org/abs/1512.06165}{{\tt
  arXiv:1512.06165}}].

\bibitem{Caldwell:2016dcw}
{\scshape MADMAX Working Group}, A.~Caldwell, G.~Dvali, B.~Majorovits,
  A.~Millar, G.~Raffelt, J.~Redondo et~al., \emph{{Dielectric Haloscopes: A New
  Way to Detect Axion Dark Matter}},
  \href{https://dx.doi.org/10.1103/PhysRevLett.118.091801}{\emph{Phys. Rev.
  Lett.} {\bf 118} (2017) 091801} [\href{https://arxiv.org/abs/1611.05865}{{\tt
  arXiv:1611.05865}}].

\bibitem{Anastassopoulos:2017ftl}
{\scshape CAST}, V.~Anastassopoulos, S.~Aune, K.~Barth, A.~Belov,
  H.~Br{\"a}uninger, G.~Cantatore et~al., \emph{New {CAST} limit on the
  axion-photon interaction},
  \href{https://dx.doi.org/10.1038/nphys4109}{\emph{Nature Phys.} {\bf 13}
  (2017) 584--590} [\href{https://arxiv.org/abs/1705.02290}{{\tt
  arXiv:1705.02290}}].

\bibitem{Baryakhtar:2018doz}
M.~Baryakhtar, J.~Huang and R.~Lasenby, \emph{{Axion and hidden photon dark
  matter detection with multilayer optical haloscopes}},
  \href{https://dx.doi.org/10.1103/PhysRevD.98.035006}{\emph{Phys. Rev. D} {\bf
  98} (2018) 035006} [\href{https://arxiv.org/abs/1803.11455}{{\tt
  arXiv:1803.11455}}].

\bibitem{armengaud2019physics}
{\scshape IAXO Collaboration}, E.~Armengaud, D.~Atti{\'e}, S.~Basso, P.~Brun,
  N.~Bykovskiy, J.~Carmona et~al., \emph{{Physics potential of the
  International Axion Observatory (IAXO)}},
  \href{https://dx.doi.org/10.1088/1475-7516/2019/06/047}{\emph{JCAP} {\bf 06}
  (2019) 047} [\href{https://arxiv.org/abs/1904.09155}{{\tt
  arXiv:1904.09155}}].

\bibitem{Lawson:2019brd}
M.~Lawson, A.~J. Millar, M.~Pancaldi, E.~Vitagliano and F.~Wilczek,
  \emph{{Tunable axion plasma haloscopes}},
  \href{https://dx.doi.org/10.1103/PhysRevLett.123.141802}{\emph{Phys. Rev.
  Lett.} {\bf 123} (2019) 141802} [\href{https://arxiv.org/abs/1904.11872}{{\tt
  arXiv:1904.11872}}].

\bibitem{gramolin2021search}
A.~V. Gramolin, D.~Aybas, D.~Johnson, J.~Adam and A.~O. Sushkov, \emph{Search
  for axion-like dark matter with ferromagnets},
  \href{https://dx.doi.org/10.1038/s41567-020-1006-6}{\emph{Nat. Phys.} {\bf
  17} (2021) 79} [\href{https://arxiv.org/abs/2003.03348}{{\tt
  arXiv:2003.03348}}].

\bibitem{Andrianavalomahefa:2020ucg}
{\scshape FUNK Experiment}, A.~Andrianavalomahefa, C.~M. Sch{\"a}fer,
  D.~Veberi{\v c}, R.~Engel, T.~Schwetz, H.-J. Mathes et~al., \emph{{Limits
  from the Funk Experiment on the Mixing Strength of Hidden-Photon Dark Matter
  in the Visible and Near-Ultraviolet Wavelength Range}},
  \href{https://dx.doi.org/10.1103/PhysRevD.102.042001}{\emph{Phys. Rev. D}
  {\bf 102} (2020) 042001} [\href{https://arxiv.org/abs/2003.13144}{{\tt
  arXiv:2003.13144}}].

\bibitem{Gelmini2020}
G.~B. Gelmini, A.~J. Millar, V.~Takhistov and E.~Vitagliano, \emph{Probing dark
  photons with plasma haloscopes},
  \href{https://dx.doi.org/10.1103/PhysRevD.102.043003}{\emph{Phys. Rev. D}
  {\bf 102} (Aug 2020) 043003} [\href{https://arxiv.org/abs/2006.06836}{{\tt
  arXiv:2006.06836}}].

\bibitem{Cantatore:2020obc}
G.~Cantatore, S.~A. Çetin, H.~Fischer, W.~Funk, M.~Karuza, A.~Kryemadhi
  et~al., \emph{Dark matter detection in the stratosphere},
  \href{https://dx.doi.org/10.3390/sym15061167}{\emph{Symmetry} {\bf 15} (2023)
  }.

\bibitem{Salemi:2021gck}
C.~P. Salemi et~al., \emph{{Search for Low-Mass Axion Dark Matter with
  ABRACADABRA-10~cm}},
  \href{https://dx.doi.org/10.1103/PhysRevLett.127.081801}{\emph{Phys. Rev.
  Lett.} {\bf 127} (2021) 081801} [\href{https://arxiv.org/abs/2102.06722}{{\tt
  arXiv:2102.06722}}].

\bibitem{Su:2021jvk}
L.~Su, L.~Wu and B.~Zhu, \emph{Probing for an ultralight dark photon from
  inverse compton-like scattering},
  \href{https://dx.doi.org/10.1103/PhysRevD.105.055021}{\emph{Phys. Rev. D}
  {\bf 105} (Mar 2022) 055021}.

\bibitem{Fedderke:2021iqw}
M.~A. Fedderke, P.~W. Graham, D.~F. {Jackson Kimball} and S.~Kalia,
  \emph{Search for dark-photon dark matter in the {SuperMAG} geomagnetic field
  dataset}, \href{https://dx.doi.org/10.1103/PhysRevD.104.095032}{\emph{Phys.
  Rev. D} {\bf 104} (2021) 095032}
  [\href{https://arxiv.org/abs/2108.08852}{{\tt arXiv:2108.08852}}].

\bibitem{Chiles_2022}
J.~Chiles, I.~Charaev, R.~Lasenby, M.~Baryakhtar, J.~Huang, A.~Roshko et~al.,
  \emph{New constraints on dark photon dark matter with superconducting
  nanowire detectors in an optical haloscope},
  \href{https://dx.doi.org/10.1103/physrevlett.128.231802}{\emph{Physical
  Review Letters} {\bf 128} (Jun 2022) }
  [\href{https://arxiv.org/abs/2110.01582}{{\tt arXiv:2110.01582}}].

\bibitem{haystaccollaboration2023new}
{\scshape {HAYSTAC} Collaboration}, M.~J. Jewell, A.~F. Leder, K.~M. Backes,
  X.~Bai, K.~van Bibber, B.~M. Brubaker et~al., \emph{New results from
  {HAYSTAC}'s phase {II} operation with a squeezed state receiver},
  \href{https://dx.doi.org/10.1103/PhysRevD.107.072007}{\emph{Phys. Rev. D}
  {\bf 107} (Apr 2023) 072007} [\href{https://arxiv.org/abs/2301.09721}{{\tt
  arXiv:2301.09721}}].

\bibitem{romanenko2023new}
A.~Romanenko, R.~Harnik, A.~Grassellino, R.~Pilipenko, Y.~Pischalnikov, Z.~Liu
  et~al., \emph{Search for dark photons with superconducting radio frequency
  cavities},
  \href{https://dx.doi.org/10.1103/PhysRevLett.130.261801}{\emph{Phys. Rev.
  Lett.} {\bf 130} (Jun 2023) 261801}
  [\href{https://arxiv.org/abs/2301.11512}{{\tt arXiv:2301.11512}}].

\bibitem{jiang2023search}
M.~Jiang, T.~Hong, D.~Hu, Y.~Chen, F.~Yang, T.~Hu et~al., \emph{Search for dark
  photons with synchronized quantum sensor network},
  \href{https://arxiv.org/abs/2305.00890}{{\tt arXiv:2305.00890}}.

\bibitem{sulai2023hunt}
I.~A. Sulai, S.~Kalia, A.~Arza, I.~M. Bloch, E.~C. Mu\~noz, C.~Fabian et~al.,
  \emph{Hunt for magnetic signatures of hidden-photon and axion dark matter in
  the wilderness},
  \href{https://dx.doi.org/10.1103/PhysRevD.108.096026}{\emph{Phys. Rev. D}
  {\bf 108} (Nov 2023) 096026}.

\bibitem{Fedderke_2021}
M.~A. Fedderke, P.~W. Graham, D.~F.~J. Kimball and S.~Kalia, \emph{Earth as a
  transducer for dark-photon dark-matter detection},
  \href{https://dx.doi.org/10.1103/physrevd.104.075023}{\emph{Physical Review
  D} {\bf 104} (Oct 2021) }.

\bibitem{goodkind}
J.~M. Goodkind, \emph{{The superconducting gravimeter}},
  \href{https://dx.doi.org/10.1063/1.1150092}{\emph{Review of Scientific
  Instruments} {\bf 70} (11 1999) 4131--4152}.

\bibitem{Griggs2017}
C.~E. Griggs, M.~V. Moody, R.~S. Norton, H.~J. Paik and K.~Venkateswara,
  \emph{Sensitive superconducting gravity gradiometer constructed with
  levitated test masses},
  \href{https://dx.doi.org/10.1103/PhysRevApplied.8.064024}{\emph{Phys. Rev.
  Appl.} {\bf 8} (Dec 2017) 064024}.

\bibitem{Hofer2023}
J.~Hofer, R.~Gross, G.~Higgins, H.~Huebl, O.~F. Kieler, R.~Kleiner et~al.,
  \emph{High-{$Q$} magnetic levitation and control of superconducting
  microspheres at millikelvin temperatures},
  \href{https://dx.doi.org/10.1103/PhysRevLett.131.043603}{\emph{Phys. Rev.
  Lett.} {\bf 131} (Jul 2023) 043603}.

\bibitem{moon2004superconducting}
F.~Moon and P.~Chang, \emph{Superconducting Levitation: Applications to Bearing
  \& Magnetic Transportation}.
\newblock Wiley VCH, 2004.

\bibitem{RomeroIsart2012}
O.~Romero-Isart, L.~Clemente, C.~Navau, A.~Sanchez and J.~I. Cirac,
  \emph{Quantum magnetomechanics with levitating superconducting microspheres},
  \href{https://dx.doi.org/10.1103/PhysRevLett.109.147205}{\emph{Phys. Rev.
  Lett.} {\bf 109} (Oct 2012) 147205}.

\bibitem{Cirio2012}
M.~Cirio, G.~K. Brennen and J.~Twamley, \emph{Quantum magnetomechanics:
  Ultrahigh-$q$-levitated mechanical oscillators},
  \href{https://dx.doi.org/10.1103/PhysRevLett.109.147206}{\emph{Phys. Rev.
  Lett.} {\bf 109} (Oct 2012) 147206}.

\bibitem{Carney_2021}
D.~Carney, A.~Hook, Z.~Liu, J.~M. Taylor and Y.~Zhao, \emph{Ultralight dark
  matter detection with mechanical quantum sensors},
  \href{https://dx.doi.org/10.1088/1367-2630/abd9e7}{\emph{New Journal of
  Physics} {\bf 23} (Mar 2021) 023041}.

\bibitem{Manley:2020mjq}
J.~Manley, M.~D. Chowdhury, D.~Grin, S.~Singh and D.~J. Wilson,
  \emph{{Searching for vector dark matter with an optomechanical
  accelerometer}},
  \href{https://dx.doi.org/10.1103/PhysRevLett.126.061301}{\emph{Phys. Rev.
  Lett.} {\bf 126} (2021) 061301} [\href{https://arxiv.org/abs/2007.04899}{{\tt
  arXiv:2007.04899}}].

\bibitem{mechanicalqs}
D.~Carney, G.~Krnjaic, D.~C. Moore, C.~A. Regal, G.~Afek, S.~Bhave et~al.,
  \emph{Mechanical quantum sensing in the search for dark matter},
  \href{https://dx.doi.org/10.1088/2058-9565/abcfcd}{\emph{Quantum Science and
  Technology} {\bf 6} (Jan 2021) 024002}.

\bibitem{Windchime:2022whs}
{\scshape Windchime}, A.~Attanasio et~al., \emph{{Snowmass 2021 White Paper:
  The Windchime Project}},  in \emph{{Snowmass 2021}}, 3 2022.
\newblock [\href{https://arxiv.org/abs/2203.07242}{{\tt arXiv:2203.07242}}].

\bibitem{Antypas:2022asj}
D.~Antypas et~al., \emph{{New Horizons: Scalar and Vector Ultralight Dark
  Matter}},  \href{https://arxiv.org/abs/2203.14915}{{\tt arXiv:2203.14915}}.

\bibitem{Li:2023wcb}
R.~Li, S.~Lin, L.~Zhang, C.~Duan, P.~Huang and J.~Du, \emph{{Search for
  Ultralight Dark Matter with a Frequency Adjustable Diamagnetic Levitated
  Sensor}}, \href{https://dx.doi.org/10.1088/0256-307X/40/6/069502}{\emph{Chin.
  Phys. Lett.} {\bf 40} (2023) 069502}
  [\href{https://arxiv.org/abs/2307.15758}{{\tt arXiv:2307.15758}}].

\bibitem{Windchime2023}
G.~Higgins, D.~Amaral and coauthors, \emph{in preparation},  (2023).

\bibitem{tinkham2004introduction}
M.~Tinkham, \emph{Introduction to Superconductivity}.
\newblock Dover Books on Physics Series. Dover Publications, 2004.

\bibitem{GutierrezLatorre2022}
M.~Gutierrez~Latorre, A.~Paradkar, D.~Hambraeus, G.~Higgins and W.~Wieczorek,
  \emph{A chip-based superconducting magnetic trap for levitating
  superconducting microparticles},
  \href{https://dx.doi.org/10.1109/TASC.2022.3147730}{\emph{IEEE Transactions
  on Applied Superconductivity} {\bf 32} (2022) 1--5}.

\bibitem{GutierrezLatorre2023}
M.~Gutierrez~Latorre, G.~Higgins, A.~Paradkar, T.~Bauch and W.~Wieczorek,
  \emph{Superconducting microsphere magnetically levitated in an anharmonic
  potential with integrated magnetic readout},
  \href{https://dx.doi.org/10.1103/PhysRevApplied.19.054047}{\emph{Phys. Rev.
  Appl.} {\bf 19} (May 2023) 054047}.

\bibitem{github}
\url{https://github.com/ZhenLiuPhys/MagLevDM}.

\bibitem{Hofer_2019}
J.~Hofer and M.~Aspelmeyer, \emph{Analytic solutions to the maxwell–london
  equations and levitation force for a superconducting sphere in a quadrupole
  field}, \href{https://dx.doi.org/10.1088/1402-4896/ab0c44}{\emph{Physica
  Scripta} {\bf 94} (Oct 2019) 125508}.

\bibitem{Schmidt2024}
P.~Schmidt, R.~Claessen, G.~Higgins, J.~Hofer, J.~J. Hansen, P.~Asenbaum
  et~al., \emph{Remote sensing of a levitated superconductor with a
  flux-tunable microwave cavity},  \href{https://arxiv.org/abs/2401.08854}{{\tt
  arXiv:2401.08854}}.

\bibitem{Chan1987a}
H.~A. Chan and H.~J. Paik, \emph{Superconducting gravity gradiometer for
  sensitive gravity measurements. {I. Theory}},
  \href{https://dx.doi.org/10.1103/PhysRevD.35.3551}{\emph{Phys. Rev. D} {\bf
  35} (Jun 1987) 3551--3571}.

\bibitem{Hansen2024}
J.~Hansen, M.~Trupke and coauthors, \emph{Readout of a levitated superconductor
  using optical interferometry}, {\emph{in preparation} (2024) }.

\bibitem{Shaw1960}
R.~W. Shaw, D.~E. Mapother and D.~C. Hopkins, \emph{Critical fields of
  superconducting tin, indium, and tantalum},
  \href{https://dx.doi.org/10.1103/PhysRev.120.88}{\emph{Phys. Rev.} {\bf 120}
  (Oct 1960) 88--91}.

\bibitem{Chanin1972}
G.~Chanin and J.~P. Torre, \emph{Critical-field curve of superconducting lead},
  \href{https://dx.doi.org/10.1103/PhysRevB.5.4357}{\emph{Phys. Rev. B} {\bf 5}
  (Jun 1972) 4357--4364}.

\bibitem{Pracht2012}
U.~S. Pracht, M.~Scheffler, M.~Dressel, D.~F. Kalok, C.~Strunk and T.~I.
  Baturina, \emph{Direct observation of the superconducting gap in a thin film
  of titanium nitride using terahertz spectroscopy},
  \href{https://dx.doi.org/10.1103/PhysRevB.86.184503}{\emph{Phys. Rev. B} {\bf
  86} (Nov 2012) 184503}.

\bibitem{Hinkle}
L.~D. Hinkle and B.~R.~F. Kendall, \emph{{Pressure‐dependent damping of a
  particle levitated in vacuum}},
  \href{https://dx.doi.org/10.1116/1.576672}{\emph{Journal of Vacuum Science \&
  Technology A} {\bf 8} (May 1990) 2802--2805}.

\bibitem{Brandt1988}
E.~H. Brandt, \emph{{Friction in levitated superconductors}},
  \href{https://dx.doi.org/10.1063/1.100435}{\emph{Applied Physics Letters}
  {\bf 53} (10 1988) 1554--1556}.

\bibitem{Arza_2022}
A.~Arza, M.~A. Fedderke, P.~W. Graham, D.~F.~J. Kimball and S.~Kalia,
  \emph{Earth as a transducer for axion dark-matter detection},
  \href{https://dx.doi.org/10.1103/physrevd.105.095007}{\emph{Physical Review
  D} {\bf 105} (May 2022) }.

\bibitem{Wilczek:1987edt}
F.~Wilczek, \emph{Two applications of axion electrodynamics},
  \href{https://dx.doi.org/10.1103/PhysRevLett.58.1799}{\emph{Phys. Rev. Lett.}
  {\bf 58} (1987) 1799--1802}.

\bibitem{RKubo_1966}
R.~Kubo, \emph{The fluctuation-dissipation theorem},
  \href{https://dx.doi.org/10.1088/0034-4885/29/1/306}{\emph{Reports on
  Progress in Physics} {\bf 29} (Jan 1966) 255}.

\bibitem{Voss1981}
R.~F. Voss, \emph{{Uncertainty principle limit to the energy sensitivity of
  SQUID’s and other linear amplifiers}},
  \href{https://dx.doi.org/10.1063/1.92271}{\emph{Applied Physics Letters} {\bf
  38} (02 1981) 182--184}.

\bibitem{Awschalom1988}
D.~D. Awschalom, J.~R. Rozen, M.~B. Ketchen, W.~J. Gallagher, A.~W.
  Kleinsasser, R.~L. Sandstrom et~al., \emph{{Low‐noise modular
  microsusceptometer using nearly quantum limited dc SQUIDs}},
  \href{https://dx.doi.org/10.1063/1.100291}{\emph{Applied Physics Letters}
  {\bf 53} (11 1988) 2108--2110}.

\bibitem{Carelli1998}
P.~Carelli, M.~G. Castellano, G.~Torrioli and R.~Leoni, \emph{{Low noise
  multiwasher superconducting interferometer}},
  \href{https://dx.doi.org/10.1063/1.121444}{\emph{Applied Physics Letters}
  {\bf 72} (01 1998) 115--117}.

\bibitem{Vasyukov2013}
D.~Vasyukov, Y.~Anahory, L.~Embon, D.~Halbertal, J.~Cuppens, L.~Neeman et~al.,
  \emph{A scanning superconducting quantum interference device with single
  electron spin sensitivity},
  \href{https://dx.doi.org/10.1038/nnano.2013.169}{\emph{Nature Nanotechnology}
  {\bf 8} (Sep 2013) 639--644}.

\bibitem{Cinquegrana1993}
C.~Cinquegrana, E.~Majorana, P.~Rapagnani and F.~Ricci,
  \emph{Back-action-evading transducing scheme for cryogenic gravitational wave
  antennas}, \href{https://dx.doi.org/10.1103/PhysRevD.48.448}{\emph{Phys. Rev.
  D} {\bf 48} (Jul 1993) 448--465}.

\bibitem{Paik2016}
H.~J. Paik, C.~E. Griggs, M.~V. Moody, K.~Venkateswara, H.~M. Lee, A.~B.
  Nielsen et~al., \emph{Low-frequency terrestrial tensor gravitational-wave
  detector},
  \href{https://dx.doi.org/10.1088/0264-9381/33/7/075003}{\emph{Classical and
  Quantum Gravity} {\bf 33} (Mar 2016) 075003}.

\bibitem{Paik1986}
H.~J. Paik, \emph{Superconducting inductance-bridge transducer for
  resonant-mass gravitational-radiation detector},
  \href{https://dx.doi.org/10.1103/PhysRevD.33.309}{\emph{Phys. Rev. D} {\bf
  33} (Jan 1986) 309--318}.

\bibitem{Budker:2013hfa}
D.~Budker, P.~W. Graham, M.~Ledbetter, S.~Rajendran and A.~Sushkov,
  \emph{{Proposal for a Cosmic Axion Spin Precession Experiment (CASPEr)}},
  \href{https://dx.doi.org/10.1103/PhysRevX.4.021030}{\emph{Phys. Rev.} {\bf
  X4} (2014) 021030} [\href{https://arxiv.org/abs/1306.6089}{{\tt
  arXiv:1306.6089}}].

\bibitem{Oharegithub}
\url{https://cajohare.github.io/AxionLimits/} (accessed 2023).

\bibitem{Caputo:2021eaa}
A.~Caputo, A.~J. Millar, C.~A. O'Hare and E.~Vitagliano, \emph{Dark photon
  limits: A handbook},
  \href{https://dx.doi.org/10.1103/physrevd.104.095029}{\emph{Physical Review
  D} {\bf 104} (Nov 2021) }.

\bibitem{Witte:2020rvb}
S.~J. Witte, S.~Rosauro-Alcaraz, S.~D. McDermott and V.~Poulin, \emph{{Dark
  photon dark matter in the presence of inhomogeneous structure}},
  \href{https://dx.doi.org/10.1007/JHEP06(2020)132}{\emph{JHEP} {\bf 06} (2020)
  132} [\href{https://arxiv.org/abs/2003.13698}{{\tt arXiv:2003.13698}}].

\bibitem{Caputo:2020bdy}
A.~Caputo, H.~Liu, S.~Mishra-Sharma and J.~T. Ruderman, \emph{Dark photon
  oscillations in our inhomogeneous universe},
  \href{https://dx.doi.org/10.1103/PhysRevLett.125.221303}{\emph{Phys. Rev.
  Lett.} {\bf 125} (2020) 221303} [\href{https://arxiv.org/abs/2002.05165}{{\tt
  arXiv:2002.05165}}].

\bibitem{escudero2023axion}
M.~Escudero, C.~K. Pooni, M.~Fairbairn, D.~Blas, X.~Du and D.~J.~E. Marsh,
  \emph{Axion star explosions: A new source for axion indirect detection},
  \href{https://arxiv.org/abs/2302.10206}{{\tt arXiv:2302.10206}}.

\bibitem{Cardoso_2018}
V.~Cardoso, {\'{O} }.~J. Dias, G.~S. Hartnett, M.~Middleton, P.~Pani and J.~E.
  Santos, \emph{Constraining the mass of dark photons and axion-like particles
  through black-hole superradiance},
  \href{https://dx.doi.org/10.1088/1475-7516/2018/03/043}{\emph{Journal of
  Cosmology and Astroparticle Physics} {\bf 2018} (Mar 2018) 043--043}.

\bibitem{bloch2023curl}
I.~M. Bloch and S.~Kalia, \emph{Curl up with a good {$\mathbf B$}: Detecting
  ultralight dark matter with differential magnetometry},
  \href{https://arxiv.org/abs/2308.10931}{{\tt arXiv:2308.10931}}.

\bibitem{Wadekar:2019xnf}
D.~Wadekar and G.~R. Farrar, \emph{{Gas-rich dwarf galaxies as a new probe of
  dark matter interactions with ordinary matter}},
  \href{https://dx.doi.org/10.1103/PhysRevD.103.123028}{\emph{Phys. Rev. D}
  {\bf 103} (2021) 123028} [\href{https://arxiv.org/abs/1903.12190v3}{{\tt
  arXiv:1903.12190v3}}].

\bibitem{McDermott:2019lch}
S.~D. McDermott and S.~J. Witte, \emph{{Cosmological evolution of light dark
  photon dark matter}},
  \href{https://dx.doi.org/10.1103/PhysRevD.101.063030}{\emph{Phys. Rev. D}
  {\bf 101} (2020) 063030} [\href{https://arxiv.org/abs/1911.05086}{{\tt
  arXiv:1911.05086}}].

\bibitem{Hoof_2023}
S.~Hoof and L.~Schulz, \emph{Updated constraints on axion-like particles from
  temporal information in supernova {SN1987A} gamma-ray data},
  \href{https://dx.doi.org/10.1088/1475-7516/2023/03/054}{\emph{Journal of
  Cosmology and Astroparticle Physics} {\bf 2023} (Mar 2023) 054}.

\bibitem{10.1093/mnras/stab3464}
J.~Sisk-Reynés, J.~H. Matthews, C.~S. Reynolds, H.~R. Russell, R.~N. Smith and
  M.~C.~D. Marsh, \emph{{New constraints on light axion-like particles using
  Chandra transmission grating spectroscopy of the powerful cluster-hosted
  quasar H1821+643}},
  \href{https://dx.doi.org/10.1093/mnras/stab3464}{\emph{Monthly Notices of the
  Royal Astronomical Society} {\bf 510} (12 2021) 1264--1277}.

\bibitem{Brady:2022bus}
A.~J. Brady, C.~Gao, R.~Harnik, Z.~Liu, Z.~Zhang and Q.~Zhuang,
  \emph{{Entangled Sensor-Networks for Dark-Matter Searches}},
  \href{https://dx.doi.org/10.1103/PRXQuantum.3.030333}{\emph{PRX Quantum} {\bf
  3} (2022) 030333} [\href{https://arxiv.org/abs/2203.05375}{{\tt
  arXiv:2203.05375}}].

\bibitem{Brady:2022qne}
A.~J. Brady et~al., \emph{{Entanglement-enhanced optomechanical sensor array
  with application to dark matter searches}},
  \href{https://dx.doi.org/10.1038/s42005-023-01357-z}{\emph{Commun. Phys.}
  {\bf 6} (2023) 237} [\href{https://arxiv.org/abs/2210.07291}{{\tt
  arXiv:2210.07291}}].

\bibitem{Xia:2022rql}
Y.~Xia, A.~R. Agrawal, C.~M. Pluchar, A.~J. Brady, Z.~Liu, Q.~Zhuang et~al.,
  \emph{{Entanglement-enhanced optomechanical sensing}},
  \href{https://dx.doi.org/10.1038/s41566-023-01178-0}{\emph{Nature Photon.}
  {\bf 17} (2023) 470--477} [\href{https://arxiv.org/abs/2210.16180}{{\tt
  arXiv:2210.16180}}].

\bibitem{Navau2021}
C.~Navau, S.~Minniberger, M.~Trupke and A.~Sanchez, \emph{Levitation of
  superconducting microrings for quantum magnetomechanics},
  \href{https://dx.doi.org/10.1103/PhysRevB.103.174436}{\emph{Phys. Rev. B}
  {\bf 103} (May 2021) 174436}.

\bibitem{GutierrezLatorre2020}
M.~G. Latorre, J.~Hofer, M.~Rudolph and W.~Wieczorek, \emph{Chip-based
  superconducting traps for levitation of micrometer-sized particles in the
  meissner state},
  \href{https://dx.doi.org/10.1088/1361-6668/aba6e1}{\emph{Superconductor
  Science and Technology} {\bf 33} (Aug 2020) 105002}.

\bibitem{purcell_morin_2013}
E.~M. Purcell and D.~J. Morin, \emph{Electricity and Magnetism}.
\newblock Cambridge University Press, {Third}~ed., 2013.

\bibitem{landau}
L.~Landau, J.~Bell, M.~Kearsley, L.~Pitaevskii, E.~Lifshitz and J.~Sykes,
  \emph{Electrodynamics of Continuous Media}.
\newblock Elsevier Science, {Second}~ed., 1984.

\bibitem{sommerfeld1959electrodynamics}
A.~Sommerfeld, \emph{Electrodynamics}.
\newblock Lectures on theoretical physics. Academic Press, 1959.

\end{thebibliography}\endgroup

\clearpage

\appendix
\onecolumngrid

\renewcommand{\theequation}{A-\arabic{equation}}
\renewcommand{\thefigure}{A-\arabic{figure}}
  \setcounter{equation}{0}
    \setcounter{figure}{0}\section{Superconducting sphere in a magnetic field}
\label{app:response}

Here, we compute the response of a superconducting sphere of radius $\mathcal R$ to an applied magnetic field of the form in \eqref{Bfield}.  Our calculation will be similar to that of \citeR[s]{Hofer_2019,Hofer2023}, but one crucial difference is that we will not assume that $\nabla\times\bm B_\mathrm{app}=0$, as this is not the case for the DPDM signal in \eqref{DPDM_signal}.  Nevertheless, we find the same result, given by \eqref{sc_force}.

As described in \secref{trapping}, the applied magnetic field causes surface currents to run on the SCP to screen the magnetic field out of its interior.  These currents then experience a Lorentz force from the (total) magnetic field, leading to a net force on the SCP.  Therefore, to calculate this force, we solve for the total magnetic field, find the corresponding surface currents, and evaluate the Lorentz force they experience.

Since the sphere is superconducting, the magnetic field inside it should vanish.%
\footnote{Here, we assume the SCP is a type-I superconductor (or a zero-field cooled type-II superconductor) so that all magnetic field lines have been expelled.  We note that physical superconductors exhibit a finite depth through which a magnetic field can penetrate into the superconductor.  This is the London penetration depth, typically $\mathcal O(10)\,\mathrm{nm}$.  So long as this depth is much smaller than $\mathcal R$, any penetration into the superconductor can be neglected.}
Let us write the total field outside the sphere as $\bm B=\bm B_\mathrm{app}+\bm B_\mathrm{resp}$, where $\bm B_\mathrm{app}$ is given by \eqref{Bfield}.  As we are interested in the instantaneous force exerted on the SCP, we may treat the magnetic field as static and the SCP as fixed.  In this case, we may choose the origin $\bm x=0$ to be the center of the sphere.  Note that this may not be the center of the trap, so that $\bm B_\mathrm{app}(\bm x=0)=\bm B_0$ may not vanish.  Let us begin by writing $\bm B_\mathrm{app}$ in terms of vector spherical harmonics (VSH).  These are defined in terms of the scalar spherical harmonics $Y_{\ell m}$ by
\begin{equation}
    \bm{Y}_{\ell m}=Y_{\ell m}\rhat,\qquad
    \bm{\Psi}_{\ell m}=r\bm{\nabla} Y_{\ell m},\qquad
    \bm{\Phi}_{\ell m}=\bm{r}\times\bm{\nabla} Y_{\ell m},
\end{equation}
(see Appendix D of \citeR{Fedderke_2021} for more details).  In terms of these VSH, we can write \eqref{Bfield} as
\begin{align}
    \bm B_\text{app}(\bm x)&=-\sqrt{\frac{2\pi}3}\left(B_{0,x}-iB_{0,y}\right)\left(\bm Y_{11}+\bm\Psi_{11}\right)+\sqrt{\frac{4\pi}3}B_{0,z}\left(\bm Y_{10}+\bm\Psi_{10}\right)+\sqrt{\frac{2\pi}3}\left(B_{0,x}+iB_{0,y}\right)\left(\bm Y_{1,-1}+\bm\Psi_{1,-1}\right)\nl
    +\sqrt{\frac\pi{30}}(b_{xx}-ib_{xy}-ib_{yx}-b_{yy})r\left(2\bm Y_{22}+\bm\Psi_{22}\right)\nl
    -\sqrt{\frac\pi{30}}(b_{xz}-ib_{yz}+b_{zx}-ib_{zy})r\left(2\bm Y_{21}+\bm\Psi_{21}\right)-\sqrt{\frac\pi6}(ib_{xz}+b_{yz}-ib_{zx}-b_{zy})r\bm\Phi_{11}\nl
    +\sqrt{\frac\pi5}b_{zz}r\left(2\bm Y_{20}+\bm\Psi_{20}\right)+\sqrt{\frac\pi3}(b_{xy}-b_{yx})r\bm\Phi_{10}\nl
    +\sqrt{\frac\pi{30}}(b_{xz}+ib_{yz}+b_{zx}+ib_{zy})r\left(2\bm Y_{2,-1}+\bm\Psi_{2,-1}\right)-\sqrt{\frac\pi6}(ib_{xz}-b_{yz}-ib_{zx}+b_{zy})r\bm\Phi_{1,-1}\nl
    +\sqrt{\frac\pi{30}}(b_{xx}+ib_{xy}+ib_{yx}-b_{yy})r\left(2\bm Y_{2,-2}+\bm\Psi_{2,-2}\right),
\label{eq:BVSH}
\end{align}
where $\bm Y_{\ell m}$, $\bm\Phi_{\ell m}$, and $\bm\Psi_{\ell m}$ are the three different types of VSH.

In the static limit, we have $\nabla\times\bm B_\mathrm{resp}=0$ and $\nabla\cdot\bm B_\mathrm{resp}=0$.  This implies that $\bm B_\mathrm{resp}$ must take the form
\begin{equation}
    \bm B_\mathrm{resp}=\sum_{n=0}^\infty a_{\ell m}r^{-\ell-2}\left(-(\ell+1)\bm Y_{\ell m}+\bm\Psi_{\ell m}\right).
\end{equation}
The radial component of the magnetic field must be continuous across the boundary of the sphere.  Since the magnetic field vanishes inside the sphere, this implies that outside the sphere, we must also have $B_r=0$.  In terms of VSH, this implies that the coefficient of the $\bm Y_{\ell m}$ modes in the total magnetic field must vanish.  This determines the coefficients $a_{\ell m}$.  It is straightforward to show then that the total field at the surface of the sphere is
\begin{align}
    \bm B(r=\mathcal R)&=-\sqrt{\frac{3\pi}2}\left(B_{0,x}-iB_{0,y}\right)\bm\Psi_{11}+\sqrt{3\pi}B_{0,z}\bm\Psi_{10}+\sqrt{\frac{3\pi}2}\left(B_{0,x}+iB_{0,y}\right)\bm\Psi_{1,-1}\nl
    +\sqrt{\frac{5\pi}{54}}(b_{xx}-ib_{xy}-ib_{yx}-b_{yy})\mathcal R\bm\Psi_{22}\nl
    -\sqrt{\frac{5\pi}{54}}(b_{xz}-ib_{yz}+b_{zx}-ib_{zy})\mathcal R\bm\Psi_{21}-\sqrt{\frac\pi6}(ib_{xz}+b_{yz}-ib_{zx}-b_{zy})\mathcal R\bm\Phi_{11}\nl
    +\sqrt{\frac{5\pi}9}b_{zz}\mathcal R\bm\Psi_{20}+\sqrt{\frac\pi3}(b_{xy}-b_{yx})\mathcal R\bm\Phi_{10}\nl
    +\sqrt{\frac{5\pi}{54}}(b_{xz}+ib_{yz}+b_{zx}+ib_{zy})\mathcal R\bm\Psi_{2,-1}-\sqrt{\frac\pi6}(ib_{xz}-b_{yz}-ib_{zx}+b_{zy})\mathcal R\bm\Phi_{1,-1}\nl
    +\sqrt{\frac{5\pi}{54}}(b_{xx}+ib_{xy}+ib_{yx}-b_{yy})\mathcal R\bm\Psi_{2,-2}.
\end{align}

The nonzero tangential magnetic field at the surface of the sphere implies some surface current $\bm K=\rhat\times\bm B$ flowing around the sphere.  The magnetic field then exerts a force on this current%
\footnote{The factor of two in \eqref{dF} originates from the discontinuity of the magnetic field at the surface of the sphere, i.e., the field just outside the sphere is $\bm B$, while the field just inside the sphere vanishes.  Heuristically, it is the average of these fields exerts a force on $\bm K$, resulting in the factor of two in \eqref{dF}.  More precisely, this factor of two can be derived by giving the magnetic field a continuous profile, which increases from 0 inside the sphere to $\bm B$ outside the sphere over some finite thickness (see Sec.~1.14 of \citeR{purcell_morin_2013} for a similar derivation force exerted by an electric field on a conductor).  \eqref{dF} can also be derived using $dF_i=\sigma_{ij}dA_j$, where $\sigma_{ij}=E_iE_j+B_iB_j-\frac12(E^2+B^2)\delta_{ij}$ is the Maxwell stress tensor~\cite{landau}.}
\begin{equation}
    \bm{dF}=\frac12(\bm K\times\bm B)dA=-\frac12B^2dA\rhat.
\label{eq:dF}
\end{equation}
Integrating this over the surface of the sphere gives
\begin{align}
    \label{eq:Fx}
    F_x&=-\frac{\mathcal R^2}2\int d\Omega\,B^2\sin\theta\cos\phi=-\pi\mathcal R^3\left((b_{xx}-b_{yy}-b_{zz})B_{0,x}+2b_{yx}B_{0,y}+2b_{zx}B_{0,z}\right)\nl
    =-\frac{3V}2\left(b_{xx}B_{0,x}+b_{yx}B_{0,y}+b_{zx}B_{0,z}\right)\\
    \label{eq:Fy}
    F_y&=-\frac{\mathcal R^2}2\int d\Omega\,B^2\sin\theta\sin\phi=-\pi\mathcal R^3\left((b_{yy}-b_{xx}-b_{zz})B_{0,y}+2b_{xy}B_{0,x}+2b_{zy}B_{0,z}\right)\nl
    =-\frac{3V}2\left(b_{xy}B_{0,x}+b_{yy}B_{0,y}+b_{zy}B_{0,z}\right)\\
    F_z&=-\frac{\mathcal R^2}2\int d\Omega\,B^2\cos\theta=-\frac{3V}2\left(b_{zz}B_{0,z}+b_{xz}B_{0,x}+b_{yz}B_{0,y}\right).
\end{align}
Note that in \eqref[s]{Fx} and (\ref{eq:Fy}), we have used the fact that $b_{xx}+b_{yy}+b_{zz}=0$ [see discussion below \eqref{Bfield}].  This total force exerted on the sphere can be expressed compactly as
\begin{equation}
    F_i=-\frac32Vb_{ji}B_{0,j}
\end{equation}
(with the Einstein summation convention implicit).  This is equivalent to \eqref{sc_force} when $\bm x$ is taken to be the center of the SCP.

\renewcommand{\theequation}{B-\arabic{equation}}
\renewcommand{\thefigure}{B-\arabic{figure}}
  \setcounter{equation}{0}
    \setcounter{figure}{0}
\section{Axion DM signal}
\label{app:axion}

In this appendix, we derive the magnetic field signal induced by axion DM in a magnetic levitation setup.  Unlike the DPDM case, axion DM requires a static magnetic field in order to convert into an observable signal.  In a magnetic levitation setup, the trap can provide this.  This is a unique feature that we exploit in this study.  Generally, this field has a complicated global configuration, and so the computation is more difficult than the DPDM case.  Here, we show how to write the result in terms of boundary integrals, which can be performed numerically.  We then determine how this result scales with the parameters of the system and evaluate it for the sample set of parameters used to compute the sensitivity in \figref{sensitivity}.

\begin{figure}[t]
\includegraphics[width=0.8\textwidth]{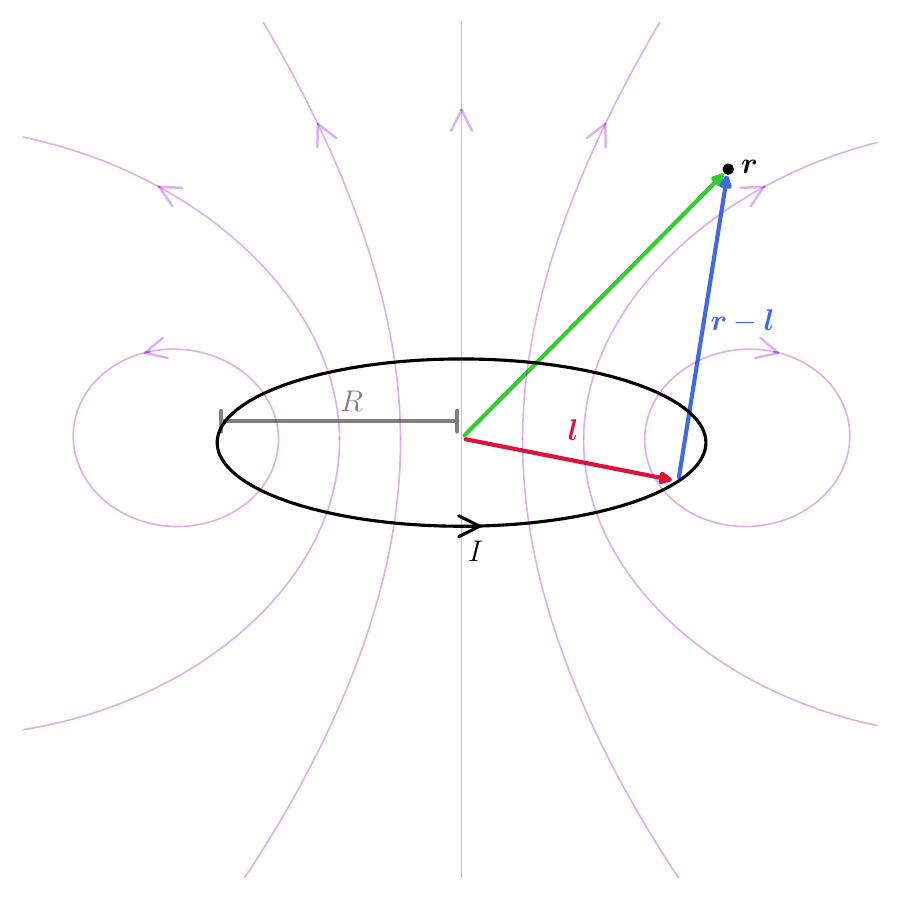}
\caption{\label{fig:loop}%
    Biot-Savart law for a single current loop of radius $R$ and current $I$.  The vector $\bm l$ (red) parametrizes the current loop.  To calculate the magnetic field at a point $\bm r$, we integrate \eqref{biot-savart} overall $\bm l$.  The relevant quantity in \eqref{biot-savart} is the distance to the loop $\bm r-\bm l$ (blue).  In purple, we show the resulting magnetic field $\bm B_\mathrm{loop}$ from a single current loop.  Note that the magnetic field is symmetric across both the $yz$- and $xz$-planes.
    }
\end{figure}

\begin{figure}[t]
\includegraphics[width=0.8\textwidth]{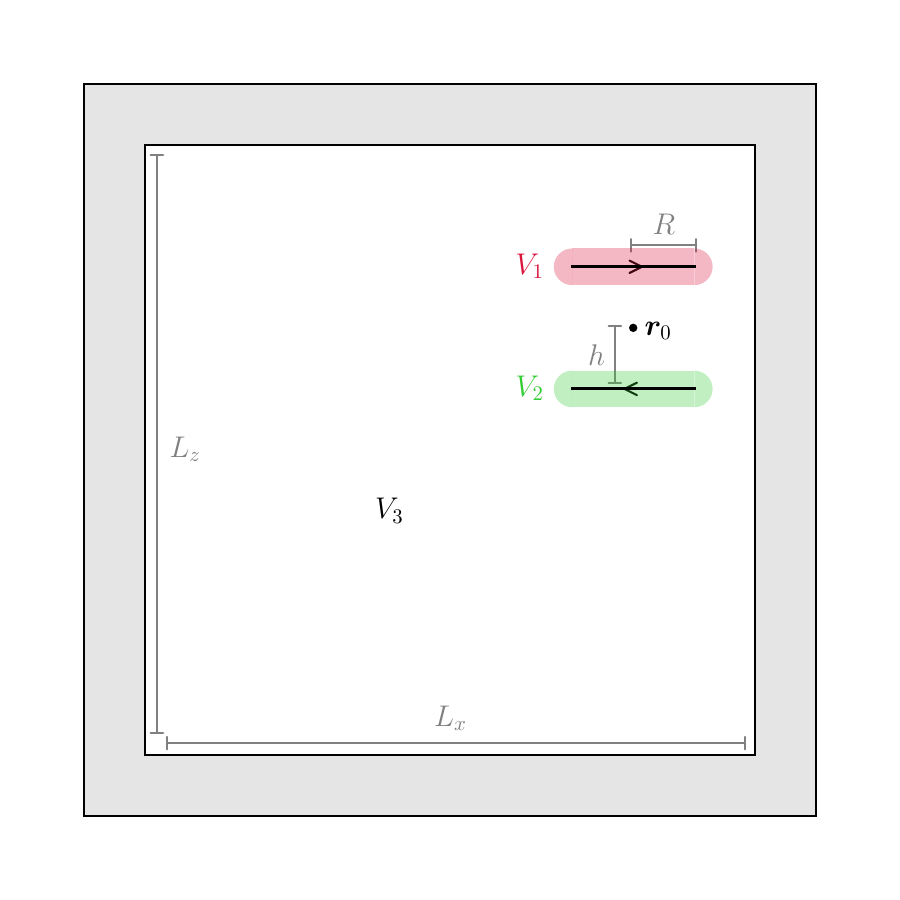}
\caption{\label{fig:shield}%
    Two-dimensional projection of levitation apparatus inside shield (not to scale).  The trap is centered at the point $\bm r_0$, and consists of two current loops, each of radius $R$.  The two loops have opposite currents and are located a distance $h$ above/below $\bm r_0$.  The trap sits inside a rectilinear shield of dimensions $L_x,L_y,L_z$.  (Note that in order to produce a nonzero signal $\bm r_0$ must be positioned off-center within the shield in \emph{at least two} directions.)  In order to evaluate the overlap integral in \eqref{cn}, we decompose the volume of the shield into three regions: $V_1$ (red) surrounding the upper loop, $V_2$ (green) surrounding the lower loop, and the rest of the volume $V_3$.}
\end{figure}

For the purposes of our calculation, we will assume the trap consists of two current loops of radius $R$ (oriented in the horizontal plane) with currents $I$ flowing in opposite directions%
\footnote{The current loops usually consist of $N$ windings, so that the total current which sources the quadrupole trap is $NI$.  For simplicity, in this section, we write $I$ in places of $NI$.}
and separated vertically by a distance $2h$ (see \figref{shield}).  By the Biot-Savart law, the magnetic field sourced by a single current loop is given by
\begin{align}
    \label{eq:biot-savart}
    \bm B_\mathrm{loop}(\bm r)&=\frac I{4\pi}\int\frac{d\bm l\times(\bm r-\bm l)}{|\bm r-\bm l|^3}\\
    &=\frac I{4\pi}\int d\theta\,\frac{Rz\cos\theta\xhat+Rz\sin\theta\yhat+(R^2-Rx\cos\theta-Ry\sin\theta)\zhat}{\left(r^2+R^2-2Rx\cos\theta-2Ry\sin\theta\right)^\frac32},
    \label{eq:Bloop}
\end{align}
where $\bm l=(R\cos\theta,R\sin\theta,0)$ parameterizes the loop and $\bm r=(x,y,z)$ is the distance to the center of the loop (see \figref{loop}).  Therefore if the trap is centered at $\bm r_0=(x_0,y_0,z_0)$, then the full magnetic field sourced by the trap is given by
\begin{equation}
    \bm B_0(\bm r)=\bm B_\mathrm{loop}(\bm r-\bm r_0-h\zhat)-\bm B_\mathrm{loop}(\bm r-\bm r_0+h\zhat).
    \label{eq:Btrap}
\end{equation}

As in \eqref{axion_current}, the effect of axion DM in the presence of this static magnetic field can be parameterized by an effective current%
\footnote{In this appendix, we promote the axion $a(\bm x,t)$ to a complex function, for calculational purposes.  The physical axion (and all resulting physical electromagnetic fields) are understood to be the real parts of the expressions given.}
\begin{equation}
    \bm J_\mathrm{eff}=ig_{a\gamma}m_aa_0\bm B_0e^{-im_at}.
\end{equation}
Then the axion DM signal can be derived by solving the Amp\`ere-Maxwell law \eqref{ampere} with this current inside our magnetic shield.  In this calculation, we will take the shield to be rectilinear, with dimensions $L_x,L_y,L_z$ (although a similar approach can be used for any geometry).  As shown in Appendix A of \citeR{Chaudhuri:2014dla}, this can be solved using a cavity mode decomposition of the shield geometry.  Let $\bm E_n$ be the electric field cavity modes of the shield (with $\bm B_n$ their associated magnetic field modes and $\omega_n$ their frequencies).  Then the magnetic field response is given by%
\footnote{For a resonant cavity, the first denominator in \eqref{cn_Jeff} would contain a damping term of the form $i\gamma_\mathrm{cav}m_a$ (with $\gamma_\mathrm{cav}\ll\omega_n$), coming from power lost to the cavity walls.  As we will be primarily concerned with the case $m_a\ll\omega_n$, this term can be safely neglected.}
\begin{equation}
    \bm B(\bm r)=\sum_nc_n\frac{\omega_n}{m_a}\bm B_n(\bm r)e^{-im_at},
\end{equation}
where
\begin{align}
    \label{eq:cn_Jeff}
    c_n&=\frac{im_a}{\omega_n^2-m_a^2}\frac{\int dV\,\bm E_n(\bm r)^*\cdot\bm J_\mathrm{eff}(t=0)}{\int dV\,|\bm E_n(\bm r)|^2}\\
    &=-\frac{g_{a\gamma}m_a^2a_0}{\omega_n^2-m_a^2}\frac{\int dV\,\bm E_n(\bm r)^*\cdot\bm B_0(\bm r)}{\int dV\,|\bm E_n(\bm r)|^2}.
    \label{eq:cn}
\end{align}
In the case of a rectilinear cavity, the two types of electric field cavity modes are TE modes (for $m,n\geq0,p\geq1$ with $m+n\neq0$) and TM modes (for $m,n\geq1,p\geq0$)
\begin{equation}
    \bm E_{\mathrm{TE},mnp}=\sqrt{\frac{2^{3-\delta_{m0}-\delta_{n0}}}{\left(\frac{m^2}{L_x^2}+\frac{n^2}{L_y^2}\right)L_xL_yL_z}}\begin{pmatrix}\frac n{L_y}\cos\left(\frac{m\pi x}{L_x}\right)\sin\left(\frac{n\pi y}{L_y}\right)\sin\left(\frac{p\pi z}{L_z}\right)\\-\frac m{L_x}\sin\left(\frac{m\pi x}{L_x}\right)\cos\left(\frac{n\pi y}{L_y}\right)\sin\left(\frac{p\pi z}{L_z}\right)\\0\end{pmatrix}
    \label{eq:TEmodes}
\end{equation}
\begin{equation}
    \bm E_{\mathrm{TM},mnp}=\sqrt{\frac{2^{3-\delta_{p0}}}{\left(\frac{m^2p^2}{L_x^2L_z^2}+\frac{n^2p^2}{L_y^2L_z^2}+\left(\frac{m^2}{L_x^2}+\frac{n^2}{L_y^2}\right)^2\right)L_xL_yL_z}}\begin{pmatrix}\frac{mp}{L_xL_z}\cos\left(\frac{m\pi x}{L_x}\right)\sin\left(\frac{n\pi y}{L_y}\right)\sin\left(\frac{p\pi z}{L_z}\right)\\\frac{np}{L_yL_z}\sin\left(\frac{m\pi x}{L_x}\right)\cos\left(\frac{n\pi y}{L_y}\right)\sin\left(\frac{p\pi z}{L_z}\right)\\-\left(\frac{m^2}{L_x^2}+\frac{n^2}{L_y^2}\right)\sin\left(\frac{m\pi x}{L_x}\right)\sin\left(\frac{n\pi y}{L_y}\right)\cos\left(\frac{p\pi z}{L_z}\right)\end{pmatrix},
    \label{eq:TMmodes}
\end{equation}
where $(0,0,0)$ denotes the bottom corner of the cavity.  (These are both normalized so that $\int dV\,|\bm E_n|^2=1$.)  These have corresponding magnetic fields
\begin{equation}
    \bm B_{\mathrm{TE},mnp}=-\frac{\pi i}{\omega_{mnp}}\sqrt{\frac{2^{3-\delta_{m0}-\delta_{n0}}}{\left(\frac{m^2}{L_x^2}+\frac{n^2}{L_y^2}\right)L_xL_yL_z}}\begin{pmatrix}\frac{mp}{L_xL_z}\sin\left(\frac{m\pi x}{L_x}\right)\cos\left(\frac{n\pi y}{L_y}\right)\cos\left(\frac{p\pi z}{L_z}\right)\\\frac{np}{L_yL_z}\cos\left(\frac{m\pi x}{L_x}\right)\sin\left(\frac{n\pi y}{L_y}\right)\cos\left(\frac{p\pi z}{L_z}\right)\\-\left(\frac{m^2}{L_x^2}+\frac{n^2}{L_y^2}\right)\cos\left(\frac{m\pi x}{L_x}\right)\cos\left(\frac{n\pi y}{L_y}\right)\sin\left(\frac{p\pi z}{L_z}\right)\end{pmatrix},
    \label{eq:BTEmodes}
\end{equation}
\begin{equation}
    \bm B_{\mathrm{TM},mnp}=\frac{i\omega_{mnp}}\pi\sqrt{\frac{2^{3-\delta_{p0}}}{\left(\frac{m^2p^2}{L_x^2L_z^2}+\frac{n^2p^2}{L_y^2L_z^2}+\left(\frac{m^2}{L_x^2}+\frac{n^2}{L_y^2}\right)^2\right)L_xL_yL_z}}\begin{pmatrix}\frac n{L_y}\sin\left(\frac{m\pi x}{L_x}\right)\cos\left(\frac{n\pi y}{L_y}\right)\cos\left(\frac{p\pi z}{L_z}\right)\\-\frac m{L_x}\cos\left(\frac{m\pi x}{L_x}\right)\sin\left(\frac{n\pi y}{L_y}\right)\cos\left(\frac{p\pi z}{L_z}\right)\\0\end{pmatrix}
    \label{eq:BTMmodes}
\end{equation}
and frequencies $\omega_{mnp}=\pi\sqrt{\frac{m^2}{L_x^2}+\frac{n^2}{L_y^2}+\frac{p^2}{L_z^2}}$.

From \eqref{cn}, we can see that the relevant quantities that we need to calculate are the overlap integrals between $\bm B_0$ and the electric field cavity modes $\bm E_n$.  Before moving to calculate these overlap integrals, let us first note some of the symmetries of $\bm B_0$, which can lead to the overlap integrals vanishing for certain positions of the trap within the rectilinear shield.  First, if we take $x\rightarrow-x$ and $\theta\rightarrow\pi-\theta$ in \eqref{Bloop}, we see that the $x$-component flips sign, but the $y$- and $z$-components do not.  This means that $\bm B_0$ is symmetric across the $yz$-plane passing through $\bm r_0$.  Note also that, for even $m$, the $y$- and $z$-components of the modes in \eqref[s]{TEmodes} and (\ref{eq:TMmodes}) flip sign under $x\rightarrow L_x-x$, while the $x$-component does not.  Therefore, if the trap is located along the central $yz$-plane of the shield (i.e., $x_0=L_x/2$), then the overlap integrals vanish for even $m$.  Moreover, from \eqref[s]{BTEmodes} and (\ref{eq:BTMmodes}), we see that the $y$- and $z$-components of $\bm B_{\mathrm{TE/TM},mnp}(x=L_x/2)$ vanish for odd $m$.  Therefore if $x_0=L_x/2$, then the axion DM signal at the center of the trap $\bm B(\bm r_0)$ must point in the $x$-direction.  A similar argument shows that if $y_0=L_y/2$, then $\bm B(\bm r_0)$ must point in the $y$-direction.  Finally, we see that taking $z\rightarrow-z$ in \eqref{Bloop} flips the signs of the $x$- and $y$-components, but not the $z$-component, and therefore taking $z\rightarrow2z_0-z$ in \eqref{Btrap} flips only the $z$-component.  In other words, $\bm B_0$ is also symmetric across the $xy$-plane passing through $\bm r_0$, and so the above argument also implies that if $z_0=L_z/2$, then $\bm B(\bm r_0)$ points in the $z$-direction.  All this means that in order to get a nonzero magnetic field signal at the center of the trap, our trap must be placed off-center within the shield in \emph{at least two} directions.

Now, let us evaluate the overlap integral.  To do so, we will decompose the volume inside the shield into three regions: two regions $V_1$ and $V_2$ which surround each current loop, and a third region $V_3$ consisting of the rest of the volume (see \figref{shield}).  Because $\nabla\times\bm B_0=0$ and $V_3$ is a simply connected region throughout which $\bm B_0$ is well-defined, then we can define a magnetic scalar potential $\Psi_0$ within $V_3$, so that $\bm B_0=\nabla\Psi_0$.  This means the overlap integral over $V_3$ simplifies into multiple boundary integrals
\begin{align}
    \int_{V_3}dV\,\bm E_n^*\cdot\bm B_0&=\int_{V_3}dV\,\bm E_n^*\cdot\nabla\Psi_0=\int_{V_3}dV\,\nabla\cdot\left(\bm E_n^*\Psi_0\right)\\
    &=\int_{\partial V}d\bm A\cdot\bm E_n^*\Psi_0-\int_{\partial V_1}d\bm A\cdot\bm E_n^*\Psi_0-\int_{\partial V_2}d\bm A\cdot\bm E_n^*\Psi_0.
    \label{eq:bound_decomp}
\end{align}
In the second equality here, we used $\nabla\cdot\bm E_n=0$.  In the final expression, the boundary $\partial V$ refers to the boundary of the shield, while the boundaries $\partial V_1$ and $\partial V_2$ refer to the boundaries of the regions $V_1$ and $V_2$.

\begin{figure}[t]
\includegraphics[width=0.8\textwidth]{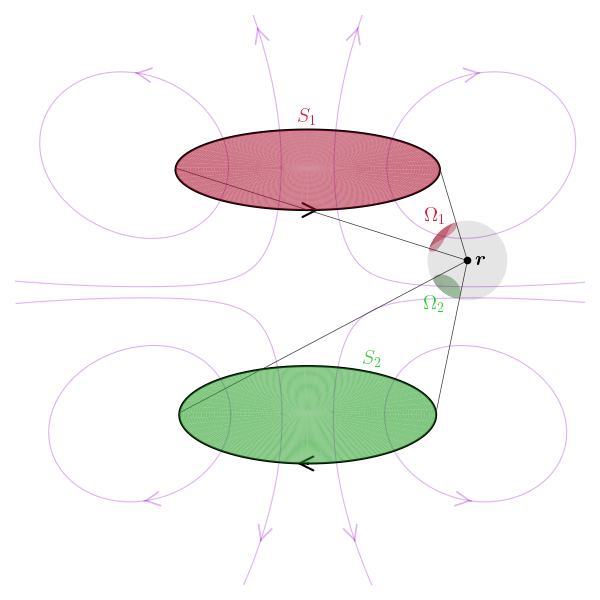}
\caption{\label{fig:omegas}%
    Magnetic field potential at a point $\bm r$.  The magnetic field potential $\bm\Psi_0(\bm r)$ defined in \eqref{potential} can be related to the solid angle subtended by each loop, as viewed from the point $\bm r$.  Specifically, we denote the surfaces bounded by the upper and lower loops by $S_1$ (red) and $S_2$ (green), respectively.  If these are projected onto a unit sphere centered at $\bm r$, they subtend angles $\Omega_1$ and $\Omega_2$, respectively, which appear in \eqref{potential}.  Note that the solid angles are defined to be positive (negative) if $\bm r$ lies above (below) the corresponding surface.  Therefore $\Omega_1<0$ and $\Omega_2>0$ here.  We also show the full magnetic field $\bm B_0$ from both loops in purple.  This magnetic field is symmetric across the $yz$-, $xz$-, and $xy$-planes.}
\end{figure}

Let $S_1$ and $S_2$ denote the surfaces bounded by the upper and lower current loops, respectively.  It is a well-known result that the magnetic scalar potential $\Psi_i(\bm r)$ from each loop individually is related to the solid angle subtended by $S_i$, as viewed from the point $\bm r$~\cite{sommerfeld1959electrodynamics}.  (This scalar potential is well-defined everywhere except on $S_i$ itself.)  Therefore, the combined potential of both loops can be written simply as
\begin{equation}
    \bm\Psi_0(\bm r)=\bm\Psi_1(\bm r)+\bm\Psi_2(\bm r)=-\frac{I(\Omega_1(\bm r)-\Omega_2(\bm r))}{4\pi},
    \label{eq:potential}
\end{equation}
where $\Omega_i$ is the solid angle subtended by $S_i$, as viewed from $\bm r$ (see \figref{omegas}).  We define $\Omega_i$ as positive if $\bm r$ lies above $S_i$, and negative otherwise.  Far from the trap $|\bm r-\bm r_0|\gg R,h$, this gives the potential of a decaying quadrupole
\begin{equation}
    \bm\Psi_0(\bm r)=-\frac I{4\pi}\left(\frac{\pi R^2(z-z_0-h)}{|\bm r-\bm r_0-h\hat z|^3}-\frac{\pi R^2(z-z_0+h)}{|\bm r-\bm r_0+h\hat z|^3}\right)=\frac{IR^2h}{2|\bm r-\bm r_0|^3}\left(1-\frac{3(z-z_0)^2}{|\bm r-\bm r_0|^2}\right).
    \label{eq:farfield}
\end{equation}
This limit can be used when evaluating the first boundary contribution in \eqref{bound_decomp}.  [Equation~(\ref{eq:farfield}) could also have been computed by taking the far-field limit of \eqref[s]{Bloop} and (\ref{eq:Btrap}).]

Next, let us consider the latter two contributions in \eqref{bound_decomp}.  We will consider the limit where we take the regions $V_1$ and $V_2$ as small as possible.  In this case, the boundary $\partial V_1$ becomes two surfaces just above and below $S_1$.  The $\Omega_2(\bm r)$ value is the same on the upper and lower surfaces of $\partial V_1$.  Therefore, the second term in \eqref{potential} cancels out when integrating over all of $\partial V_1$.  However, $\Omega_1(\bm r)$ approaches $2\pi$ as $\bm r$ approaches $S_1$ from above, while it approaches $-2\pi$ as $\bm r$ approaches from below, so the former term in \eqref{potential} does not cancel!  (Recall that $\Psi_1$ is not defined on $S_1$, and so it can exhibit a discontinuity there.)  We instead find
\begin{equation}
    \int_{\partial V_1}d\bm A\cdot\bm E_n^*\Psi_0=-I\int_{S_1}d\bm A\cdot\bm E_n^*.
\end{equation}
A similar statement applies for the boundary integral over $\partial V_2$, but with the opposite sign.

\begin{figure}[t]
\includegraphics[width=0.8\textwidth]{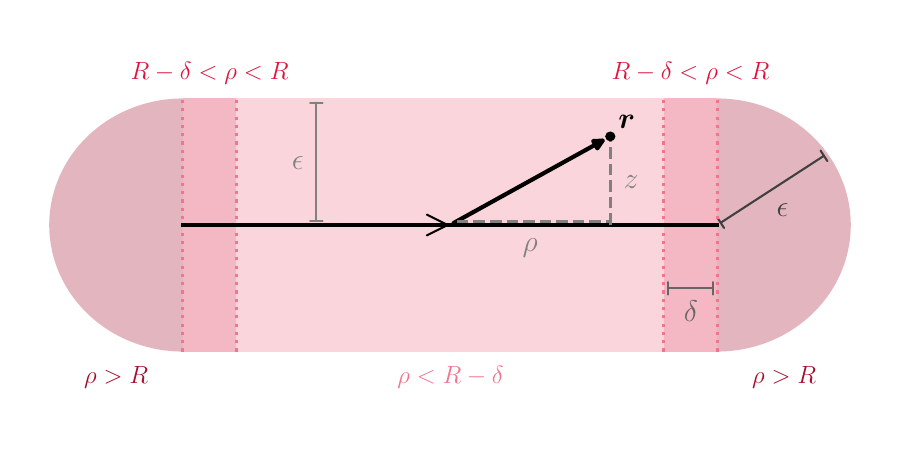}
\caption{\label{fig:v1int}%
    Decomposition of $V_1$.  We define $V_1$ as the set of points which are within a distance $\epsilon$ of the surface $S_1$, and parametrize it using cylindrical coordinates $\bm r=(\rho,\theta,z)$.  In order to show that the overlap integral over $V_1$ vanishes as $\epsilon\rightarrow0$, we decompose it into three regions: $\rho>R$ (dark red), $R-\delta<\rho<R$ (medium red), and $\rho<R-\delta$ (light red), for some fixed $\delta\ll R$.  The magnetic field in the light red region remains finite as we take $\epsilon\rightarrow0$.  Meanwhile, the volume of the dark red region shrinks faster than its magnetic field grows, as $\epsilon\rightarrow0$.  Therefore, these contributions vanish trivially.  Equations~(\ref{eq:reg2_first})--(\ref{eq:reg2_last}) show that the contribution from the medium red region also vanishes.
    }
\end{figure}

Now that we know how to evaluate the overlap integral over $V_3$, let us consider the contributions from $V_1$ and $V_2$.  We will find that these vanish in the limit that $V_1$ and $V_2$ approach the surfaces $S_1$ and $S_2$.  This will be due to the fact that the volumes of $V_1$ and $V_2$ vanish in this limit.  However, we should consider this limit carefully, as $\bm B_0$ (and therefore the integrands) also diverge near the current loops.  For concreteness, let us define $V_1$ as the set of points within a distance $\epsilon$ of $S_1$, and take the limit as $\epsilon\rightarrow0$.  Let us define $\rho^2=x^2+y^2$, and then separate $V_3$ into three regions: $\rho<R-\delta$, $R-\delta<\rho<R$, and $\rho>R$, for some $\delta\ll R$ which remains fixed as we take the limit $\epsilon\rightarrow0$ (see \figref{v1int}).  The integrand does not diverge in the first region, so the integral over this region vanishes trivially as $\epsilon\rightarrow0$.  The integrand in the third region does diverge as $\bm B_0\propto1/\epsilon$; however, the volume of the region goes as $\epsilon^2$.  Therefore, the contribution from the third region should vanish as well.  The second region requires more careful treatment.  In the limit of small $\delta$, we can approximate $\bm B_0$ as the magnetic field from a straight wire.  Then, explicitly, the integral over the second region looks like
\begin{align}
    \label{eq:reg2_first}
    \int_0^{2\pi}d\theta\int_{-\epsilon}^\epsilon dz\int_{R-\delta}^Rd\rho\,&\bm E_n^*\cdot\left(\frac I{2\pi\sqrt{z^2+(R-\rho)^2}}\cdot\frac{(R-\rho)\zhat+z\rhohat}{\sqrt{z^2+(R-\rho)^2}}\right)\\
    &=\int_0^{2\pi}d\theta~\frac{I\bm E_n^*}{2\pi}\cdot\int_{-\epsilon}^\epsilon dz\int_0^\delta d\tilde\rho~\frac{\tilde\rho\zhat+z\rhohat}{z^2+\tilde\rho^2}\\
    &=\int_0^{2\pi}d\theta~\frac{I\bm E_n^*}{2\pi}\cdot\int_{-\epsilon}^\epsilon dz~\frac12\log\left(\frac{z^2+\delta^2}{z^2}\right)\zhat\\
    &=\int_0^{2\pi}d\theta~\frac{I\bm E_n^*}{2\pi}\cdot\left(\epsilon\log\left(\frac{\epsilon^2+\delta^2}{\epsilon^2}\right)-2\delta\tan^{-1}\frac\delta\epsilon+\pi\delta\right)\zhat,
    \label{eq:reg2_last}
\end{align}
which vanishes in the limit $\epsilon\rightarrow0$.  Therefore, the volume integral from $V_1$ does not contribute to the full overlap integral in \eqref{cn}.  A similar argument shows that the volume integral from $V_2$ does not contribute either.

In summary, we find that the full overlap integral appearing in \eqref{cn} is the sum of the contributions in \eqref{bound_decomp}.  We can write the full solution for the magnetic field signal at the center of the trap as
\begin{equation}
    \bm B(\bm r_0)=-g_{a\gamma}m_aa_0e^{-im_at}\sum_n\frac1{\omega_n}\left(\int_{\partial V}d\bm A\cdot\bm E_n^*\Psi_0+I\int_{S_1}d\bm A\cdot\bm E_n^*-I\int_{S_2}d\bm A\cdot\bm E_n^*\right)\bm B_n(\bm r_0).
    \label{eq:signal_decomp}
\end{equation}
If $L_x,L_y,L_z\gg R,h$, then $\Psi_0$ in the first integral can be evaluated using \eqref{potential}.  We have assumed here that $m_a\ll\omega_n$ and that the modes are normalized so that $\int dV\,|\bm E_n|^2=1$.  Generically, \eqref{signal_decomp} needs to be evaluated numerically, but we can determine a parametric estimate analytically.  The normalization of $\bm E_n$ fixes $\bm E_n\sim L^{-3/2}$ [see \eqref[s]{TEmodes} and (\ref{eq:TMmodes})], where $L_x,L_y,L_z\sim L$, so from \eqref{potential}, we find that
\begin{equation}
    \int_{\partial V}d\bm A\cdot\bm E_n^*\Psi_0\sim\frac{IR^2h}{L^{5/2}}.
\end{equation}
Note that when $\omega_n h\gg1$, the integrals over $S_1$ and $S_2$ are nearly equal.  In this case, we can Taylor expand $\bm E_n^*$ around $\bm r_0$ to rewrite their difference as
\begin{equation}
    I\int_{S_1}d\bm A\cdot\bm E_n^*-I\int_{S_2}d\bm A\cdot\bm E_n^*\approx2\pi IR^2h\cdot\partial_zE_{n,z}(\bm r_0)^*\sim\frac{IR^2h}{L^{5/2}},
\end{equation}
so the boundary integral contributions scale similarly with the parameters of the system.  In order to connect with the magnetic field gradient $b$ introduced in \eqref{Bfield}, let us note that the applied magnetic field near the center of the trap $\bm r_0$ is
\begin{equation}
    \bm B_0(\bm r)=-\frac{3IR^2h}{\left(R^2+h^2\right)^{5/2}}\left(\frac{x-x_0}2,\frac{y-y_0}2,-z+z_0\right)\equiv b_0\left(\frac{x-x_0}2,\frac{y-y_0}2,-z+z_0\right).
    \label{eq:trap_center}
\end{equation}
Then from \eqref{signal_decomp}, we find that the axion DM magnetic field signal scales as
\begin{equation}
    \bm B(\bm r_0)\sim\frac{g_{a\gamma}m_aa_0b_0\left(R^2+h^2\right)^{5/2}}{L^3}.
\end{equation}
The exact constant of proportionality will depend on the position of the trap within the shield.  (Recall that for a sufficiently symmetric position, the constant may be zero.)  For instance, for $L_x=L_y=L_z=10\,\mathrm{cm}$, $R=h=1\,\mathrm{cm}$, and $\bm r_0=(7\,\mathrm{cm},8\,\mathrm{cm},5\,\mathrm{cm})$ [which are the parameters used for the blue curves in \figref{sensitivity}],%
\footnote{Numerically, we find that the $z$-component of $\bm B(\bm r_0)$ tends to converge faster than the $x$- and $y$-components.  Therefore we take $z_0=L_z/2$ here, so the magnetic field signal points exactly in the $z$-direction, and we can calculate the signal accurately.}
we find the constant of proportionality to be roughly 0.09.

\end{document}